\documentclass[12pt, a4paper]{article}
\pdfoutput=1
\usepackage[dvipdfmx]{graphicx}
\usepackage{amssymb}
\usepackage{amsmath}
\usepackage{bm}
\usepackage{color}
\usepackage{cite}
\usepackage{slashed}
\usepackage{subfigure}
\usepackage{epstopdf}            
\usepackage{epsfig}
\newcommand{\beq}{\begin{equation}}
\newcommand{\eeq}{\end{equation}}

\usepackage[colorlinks=true, linkcolor=black, citecolor=black,
urlcolor=blue]{hyperref} 

\begin{document}

\begin{titlepage}
\begin{center}

%\Large{Testing the new physics interpretation of the muon g-2 anomaly at the LHC}}
%{\color{red}
{\Large Testing the 2HDM explanation of the muon g-2 anomaly
\\ at the LHC}
%}
%%%%%%%%%%%%%%%%%%%%%%%

\vskip 2cm
Syuhei Iguro$^1$, Yuji Omura$^{2}$ and Michihisa Takeuchi$^{3}$
\vskip 0.5cm

{\it $^1$
Department of Physics, Nagoya University, Nagoya 464-8602, Japan}\\[3pt]

{\it $^2$ 
Department of Physics, Kindai University, Higashi-Osaka, Osaka 577-8502, Japan}\\[3pt]

{\it $^3$
Kobayashi-Maskawa Institute for the Origin of Particles and the
Universe, \\ Nagoya University, Nagoya 464-8602, Japan}\\[3pt]

%%%%%%%%%%%%%%
\vskip 1.5cm

\begin{abstract}
The discrepancy between the measured value and the
Standard Model prediction for the muon anomalous magnetic moment is one of the important issues 
in the particle physics. In this paper, we consider a two Higgs doublet model (2HDM) where the extra Higgs doublet 
couples to muon and tau in lepton flavor violating (LFV) way and the one-loop correction involving the 
scalars largely contributes to the muon anomalous magnetic moment. 
The couplings should be sizable to explain the discrepancy, so that the extra Higgs bosons would dominantly 
decay into $\mu\tau$ LFV modes, which makes the model testable at the LHC through multi-lepton signatures 
even though they are produced via the electroweak interaction.
We discuss the current status and the future prospect for the extra Higgs searches at the LHC, and  
demonstrate the reconstruction of the mass spectrum using the multi-lepton events.
\end{abstract}
\end{center}
\end{titlepage}

%%%%%%%%%%%%%%%%%%%%%%%%%%%%%%%%%%%%
%%%%%%%%%%%%%%%%%%%%%%%%%%%%%%%%%%%%
\section{Introduction}

In recent years, we can perform high precision verification of the Standard Model (SM).
Many physical observables have been measured with the high accuracy and their SM 
predictions have been also well developed. We can test not only the SM but also new 
physics beyond the SM, comparing the theoretical predictions with the experimental results.
Most of the results suggest that the SM describes our nature very well, 
while we also find some measurements deviated from the SM predictions. 
One of the well-known observables that shows a discrepancy is the anomalous magnetic moment of muon.

Taking the quantum corrections into account, the magnetic moment is deviated from two, 
and the deviation is called the anomalous magnetic moment. 
The muon anomalous magnetic moment is usually denoted as $a_\mu=(g-2)_\mu/2$ and 
measured with the fairly high accuracy. The latest experimental result is given by E821 experiment at the Brookhaven National Lab (BNL) as $a_\mu^{\rm exp}=11 659 208.0(5.4)(3.3) \times10^{-10}$~\cite{Bennett:2006fi}. The new experiments at the Fermilab (FNAL)~\cite{Grange:2015fou} and at the J-PARC~\cite{Mibe:2011zz} are scheduled, and they will measure it more precisely. On the other hand, the SM prediction that takes into account the higher-loop correction
involving the heavy fermions as well as the gauge bosons is given so far by several groups~\cite{Hagiwara:2006jt,Jegerlehner:2009ry,Davier:2010nc,Hagiwara:2011af},
and there is a consistent deviation between the measured value $a_{\mu}^{\rm{exp}}$ and the SM prediction $a_{\mu}^{\rm{SM}}$ at a $3$-$4 ~\sigma$ level. In this paper, we take the following value reported in Ref.~\cite{Hagiwara:2006jt} as a nominal value of the deviation,
\begin{align}
\delta a_\mu^{\rm exp} = a_{\mu}^{\rm exp} - a_{\mu}^{\rm{SM}} = (27.8 \pm 8.2) \times 10^{-10}.
\end{align}
This fact indicates a possibility of existence of unknown new particles in the loop, so that this measurement plays an important role in searching for new physics beyond the SM.
\medskip

Motivated by the discrepancy, many new physics interpretations have been proposed.
One of the simplest models is a 2HDM, where an extra Higgs doublet is introduced to the SM. 
If there is no symmetry to distinguish the two Higgs doublets,
both Higgs fields couple to the SM fermions. 
In general, the extra scalars that appear after the electroweak (EW) symmetry breaking can have
flavor-dependent couplings to quarks and leptons at the tree level.
If the flavor violating couplings involving $\mu$ and $\tau$ are sizable with appropriate signs, we can simply enhance $a_\mu$~\cite{Omura:2015nja,Omura:2015xcg,Iguro:2018qzf,Crivellin:2019dun}. 
\footnote{There are other possible setups to explain this anomaly:
a muon specific 2HDM \cite{Abe:2017jqo}, a lepton specific (Type-X) 2HDM \cite{Broggio:2014mna,Wang:2014sda,Abe:2015oca,Chun:2015hsa,Chun:2016hzs}, a aligned 2HDM \cite{Ilisie:2015tra,Han:2015yys,Cherchiglia:2016eui,Cherchiglia:2017uwv}, a $U(1)$-symmetric 2HDM \cite{Li:2018aov} and a perturbed 2HDM \cite{Crivellin:2015hha}.  For a recent review, see also \cite{Lindner:2016bgg}.}
This setup is, in fact, very successful in explaining the anomaly in $a_\mu$ and at the same time in evading the strict experimental constraints from flavor physics, although many tree-level flavor changing neutral currents (FCNCs) involving scalars are assumed to be suppressed by hand or by some mechanisms. Such a 2HDM with tree-level FCNCs is obtained as the effective field theory of 
the Left Right symmetric models~\cite{Iguro:2018oou}, the variant axion models~\cite{Chiang:2015cba,Chiang:2017fjr,Chiang:2018bnu}, leptoquark models~\cite{IKOO} and so on.
Moreover, it is recently pointed out that this unique alignment of the scalar couplings required to accomodate the anomaly can be realized by a specific $Z_4$ flavor symmetry~\cite{Abe:2019bkf}.
\medskip

One important issue relevant to this setup is how to probe this scenario in experiments. As discussed in Ref. \cite{Omura:2015xcg}, various flavor processes severely constrain the scalar masses and the $\mu\tau$ couplings, if the other couplings are also
sizable. On the other hand, it turns out to be difficult to test it
when the $\mu\tau$ couplings dominate over the other Yukawa elements. 
Our main purpose of this paper is to point out the new distinctive signatures at the LHC, which is conventionally uncovered.
\footnote{In Ref.~\cite{Altmannshofer:2016brv}, the similar signatures has 
been studied in the model with an extra gauge boson.}

As we will show, the masses of the extra Higgs scalars are required to be ${\cal O}(100)$ GeV 
to explain the deviation, and the $\mu\tau$ couplings are expected to be ${\cal O}(0.1-1)$.
In this case, these Higgs scalars can be produced in pair through the EW interaction at the 
LHC with a visible rate. The neutral ones decay into $\mu\tau$, and the charged one does 
into $\mu$ and $\nu_\tau$ or $\tau$ and $\nu_\mu$, therefore, especially from the two neutral 
scalar pair-production ($HA$), $\mu^+ \mu^- \tau^+ \tau^-$ and $\mu^\pm \mu^\pm \tau^\mp \tau^\mp$ 
signatures are expected, and the latter same-sign di-muon with same-sign di-tau is a very 
distinctive signature. The other production modes also contribute to the multi-lepton final states.
When one charged scalar and one neutral scalar are produced in pair, $\mu\mu\tau \nu$ and 
$\mu\tau\tau\nu$ final states are expected. When two charged scalars are produced, $\mu^+\mu^-$, 
$\tau^+\tau^-$, and $\mu^\pm\tau^\mp$ signals, associated with two neutrinos are expected in the final state. 
We study the signals induced by the heavy Higgs pair production, and summarize the current 
status and the future prospects of this model. In particular, the signatures with the same-sign 
muons and the same-sign taus play a crucial role.
\medskip

This paper is organized as follows. In Sec. 2, we briefly introduce our setup in the 2HDM 
and discuss the relevant parameter space for explaining the deviation of $a_\mu$.
In Sec. 3, we discuss the collider phenomenology in detail and show how we can determine 
the mass spectrum from the multi-lepton events. Sec. 4 is devoted to the discussion of the 
other miscellaneous issues, and the summary is given in Sec. 5. 

%%%%%%%%%%%%%%%%%%%%%%%%%%%%%%%%%%%%
%%%%%%%%%%%%%%%%%%%%%%%%%%%%%%%%%%%%
\section{Setup and the contribution to $a_\mu$}
\subsection{2HDM with tree-level FCNCs}
\label{Model}
We consider the extended SM with an extra Higgs doublet. In general, both Higgs doublets 
develop non-vanishing vacuum expectation values (VEVs), but we can choose the basis into the so-called 
Higgs basis~\cite{Georgi:1978ri,Donoghue:1978cj}, where only one Higgs doublet has 
the non-vanishing VEV. In this basis, the two Higgs doublets can be decomposed as
\begin{eqnarray}
  H_1 =\left(
  \begin{array}{c}
    G^+\\
    \frac{v+\phi_1+iG}{\sqrt{2}}
  \end{array}
  \right),~~~
  H_2=\left(
  \begin{array}{c}
    H^+\\
    \frac{\phi_2+iA}{\sqrt{2}}
  \end{array}
  \right),
\label{HiggsBasis}
\end{eqnarray}
where $G^+$ and $G$ are Nambu-Goldstone bosons, and $H^+$ and $A$ are a charged Higgs boson and a CP-odd
Higgs boson, respectively. Note that $H_1$ corresponds to the Higgs field with the non-vanishing VEV, denoted as $v/\sqrt 2\simeq174$ GeV.
The CP-even neutral Higgs bosons, $\phi_1$ and $\phi_2$, mix each other and form the mass
eigenstates, $h$ and $H$. We identify $h$ as the Higgs boson with $125$ GeV mass and assume $m_H>m_h$ in this paper. The mixing angle $\theta_{\beta \alpha}$ is conventionally described as
\begin{eqnarray}
  \left(
  \begin{array}{c}
    \phi_1\\
    \phi_2
  \end{array}
  \right)=\left(
  \begin{array}{cc}
    \cos\theta_{\beta \alpha} & \sin\theta_{\beta \alpha}\\
    -\sin\theta_{\beta \alpha} & \cos\theta_{\beta \alpha}
  \end{array}
  \right)\left(
  \begin{array}{c}
    H\\
    h
  \end{array}
  \right).
\end{eqnarray}
In the limit of vanishing $\cos\theta_{\beta \alpha}$   ($\sin\theta_{\beta\alpha}\rightarrow 1$), 
the interaction of $h$ becomes identical to the one in the SM. If any discrete symmetry is 
not imposed, both Higgs doublets can couple to all fermions. In the mass eigenstates 
of the fermions, the Yukawa interactions are expressed as
\begin{eqnarray}
  {\cal L} &=&-\bar{Q}_L^i H_1 y_d^i d_R^i -\bar{Q}_L^i H_2 \rho_d^{ij} d_R^j-\bar{Q}_L^i (V^\dagger)^{ij} \tilde{H}_1 y_u^j u_R^j
  -\bar{Q}_L^i (V^\dagger)^{ij} \tilde{H}_2 \rho_u^{jk}u_R^k \nonumber\\
  &&-\bar{L}_L^i H_1 y^i_e e_R^i -\bar{L}_L^i H_2 \rho^{ij}_e e_R^j+{\rm h.c.}.
\label{yukawas}
\end{eqnarray}
Here $i$ and $j$ represent the flavor indices, and $\tilde{H}_{1,2}=i \sigma_2 H^*_{1,2}$ is defined using the Pauli matrix, $\sigma_2$. The left-handed fermions are defined as $Q_L=(V^\dagger u_L,d_L)^T$ and $L=(V_{\rm MNS} \nu_L, e_L)^T$, where $V$ and $V_{\rm MNS}$ are the Cabbibo-Kobayashi-Maskawa (CKM) and the Maki-Nakagawa-Sakata (MNS) matrices, respectively. 
Note that Yukawa couplings $y_f^i$ are defined as $y_f^i=\sqrt{2}m_f^i/v$ using the fermion masses $m_f^i$.
The Yukawa couplings, $\rho_f^{ij}$, are, on the other hand, unknown general $3\times3$ complex matrices and are the sources of the Higgs-mediated flavor violation.

In the mass eigenstates of the Higgs bosons, the Yukawa interactions are given by
\begin{align}
\label{YukawaL}
  {\cal L}&=-\sum_{f=u,d,e}\sum_{\phi=h,H,A} y^f_{\phi i j}\bar{f}_{Li} \phi f_{Rj}+{\rm h.c.}
  \nonumber\\
  &\quad -\bar{\nu}_{Li} (V_{\rm MNS}^\dagger \rho_e)^{ij}  H^+ e_{Rj}
  -\bar{u}_i(V\rho_d P_R-\rho_u^\dagger V P_L)^{ij} H^+d_j+{\rm h.c.},
\end{align}
where
\begin{align}
  y^f_{hij}=\frac{m_{f}^i}{v}s_{\beta\alpha}\delta_{ij}&+\frac{\rho_{f}^{ij}}{\sqrt{2}} c_{\beta\alpha},~~
  y^f_{Hij}=\frac{m_f^i}{v} c_{\beta \alpha}\delta_{ij}-\frac{\rho_f^{ij}}{\sqrt{2}} s_{\beta\alpha},
  \nonumber \\
  y^f_{Aij}&=
  \left\{
  \begin{array}{c}
    -\frac{i\rho_f^{ij}}{\sqrt{2}}~({\rm for}~f=u),\\
    \frac{i\rho_f^{ij}}{\sqrt{2}}~({\rm for}~f=d,~e),
  \end{array}
  \right.
  \label{yukawa}
\end{align}
and $c_{\beta\alpha}$ and $s_{\beta\alpha}$ are short for $\cos\theta_{\beta\alpha}$ and $\sin\theta_{\beta\alpha}$ respectively.
We note that when $c_{\beta\alpha}$ is vanishing, the Yukawa interaction of $h$ becomes identical to the one in the SM. Throughout this paper, we simply assume that $c_{\beta\alpha}=0$ to avoid the constraints on the 125-GeV Higgs particle.

While the Yukawa interactions of heavy Higgs bosons ($H$, $A$, and $H^+$)
are controlled by the $\rho_f^{ij}$ couplings, 
the mass of the heavy scalars are controlled by the Higgs potential, 
$V(H_i)=\lambda_4 (H_1^\dagger H_2)(H_2^\dagger H_1) +\{ \frac{\lambda_5}{2}(H_1^\dagger H_2)^2 +{\rm h.c.}\}+\cdots$.
In particular, their mass differences are given by the dimensionless parameters in the Higgs potential as follows:
\begin{eqnarray}
  m_H^2 \simeq m_A^2+\lambda_5 v^2, ~~~~~~
  m_{H^\pm}^2 \simeq m_A^2-\frac{\lambda_4-\lambda_5}{2} v^2,
  \label{Higgs_spectrum}
\end{eqnarray}
where $m_H$, $m_A$ and $m_{H^\pm}$ denote the masses of the heavy CP-even, the CP-odd, 
and the charged Higgs scalars, respectively.
We consider the case that sizable $\rho_{e}^{\mu\tau}$ and $\rho_{e}^{\tau\mu}$ induce enough contribution to $\delta a_{\mu}$.
Once we turn them on, other Yukawa elements are strictly constrained by the flavor and the collider physics~\cite{Omura:2015xcg,Iguro:2018qzf}.
For example, $\rho_u^{tt}$ should be small since a non-vanishing $\rho_u^{tt}$ with sizable $\rho_{e}^{\mu\tau}$ and $\rho_{e}^{\tau\mu}$ 
devastatingly enhances the $BR(\tau\to\mu\gamma)$.
Therefore, phenomenologically, we consider the situation that $\rho_{e}^{\mu\tau}$ and $\rho_{e}^{\tau\mu}$ are only sizable, 
and the other Yukawa elements are negligibly small.
We give a comment on the contribution of the other elements in Sec.~\ref{DCS}.
Note that we consider $0\le\lambda_5\le 1$ to avoid the unstable vacuum and the non-perturbative couplings in our paper.
 
%%%%%%%%%%%%%%%%%%%%%
\subsection{parameter region to explain $\delta a_\mu$}
\label{muong2}
In our scenario of the 2HDM, the sizable contribution to $\delta a_\mu$ is generated 
via the 1-loop diagram mediated by the extra neutral Higgs bosons $H$ and $A$, and is given as \cite{Omura:2015nja,Omura:2015xcg}
\begin{align}
\delta a_\mu&=a_{\mu}^{\rm{2HDM}}-a_{\mu}^{\rm{SM}} \nonumber\\
&=\frac{m_\mu m_\tau\rho_{e}^{\mu\tau}\rho_{e}^{\tau\mu}}{16\pi^2}\left( \frac{\ln{\frac{m_H^2}{m_\tau^2}}-\frac{3}{2}}{m_H^2}-\frac{\ln{\frac{m_A^2}{m_\tau^2}}-\frac{3}{2}}{m_A^2}\right)\nonumber\\
&\simeq-\frac{m_\mu m_\tau\rho_{e}^{\mu\tau}\rho_{e}^{\tau\mu}}{8\pi^2}\frac{\Delta_{H-A}}{m_A^3}\left(\ln{\frac{m_A^2}{m_\tau^2}}-\frac{5}{2} \right)\nonumber\\
&\simeq -3\times10^{-9}\left(\frac{\rho_{e}^{\mu\tau}\rho_{e}^{\tau\mu}}{0.3^2}\right)\left(\frac{\Delta_{H-A}}{60\rm{[GeV]}}\right) \left(\frac{300[\rm{GeV}]}{m_A}\right)^3,
\label{muon_g-2}
\end{align}
where the mass difference between $H$ and $A$ is denoted as $\Delta_{H-A}=m_H-m_A$ and $a_{\mu,\rm{2HDM}}$ is the prediction of $a_\mu$ in the 2HDM.
We find that the parameters relevant to $\delta a_\mu$ are $\rho_{e}^{\mu\tau} \rho_{e}^{\tau\mu}$, $m_A$ and $m_H$, and the product $\Delta_{H-A}\rho_{e}^{\mu\tau} \rho_{e}^{\tau\mu}$ must be negative
to obtain the positive value.\footnote{The contribution from the $H^\pm$-loop diagram does not have a $m_\tau$ enhancement and is small. }
Note that $\lambda_5 \ge 0$ implies $\Delta_{H-A} \ge 0$  from Eq.(\ref{Higgs_spectrum}), and
$\delta a_\mu$ vanishes when $\Delta_{H-A}=0$.
The mass of the charged scalar $m_{H^\pm}$ must be sufficiently degenerated 
with either of $m_H$ or $m_A$ to evade the stringent constraint 
from the electroweak precision tests \cite{Peskin:1991sw}
although not directly relevant to the $\delta a_\mu$.

\begin{figure}[h]
  \begin{center}
    \includegraphics[width=10cm]{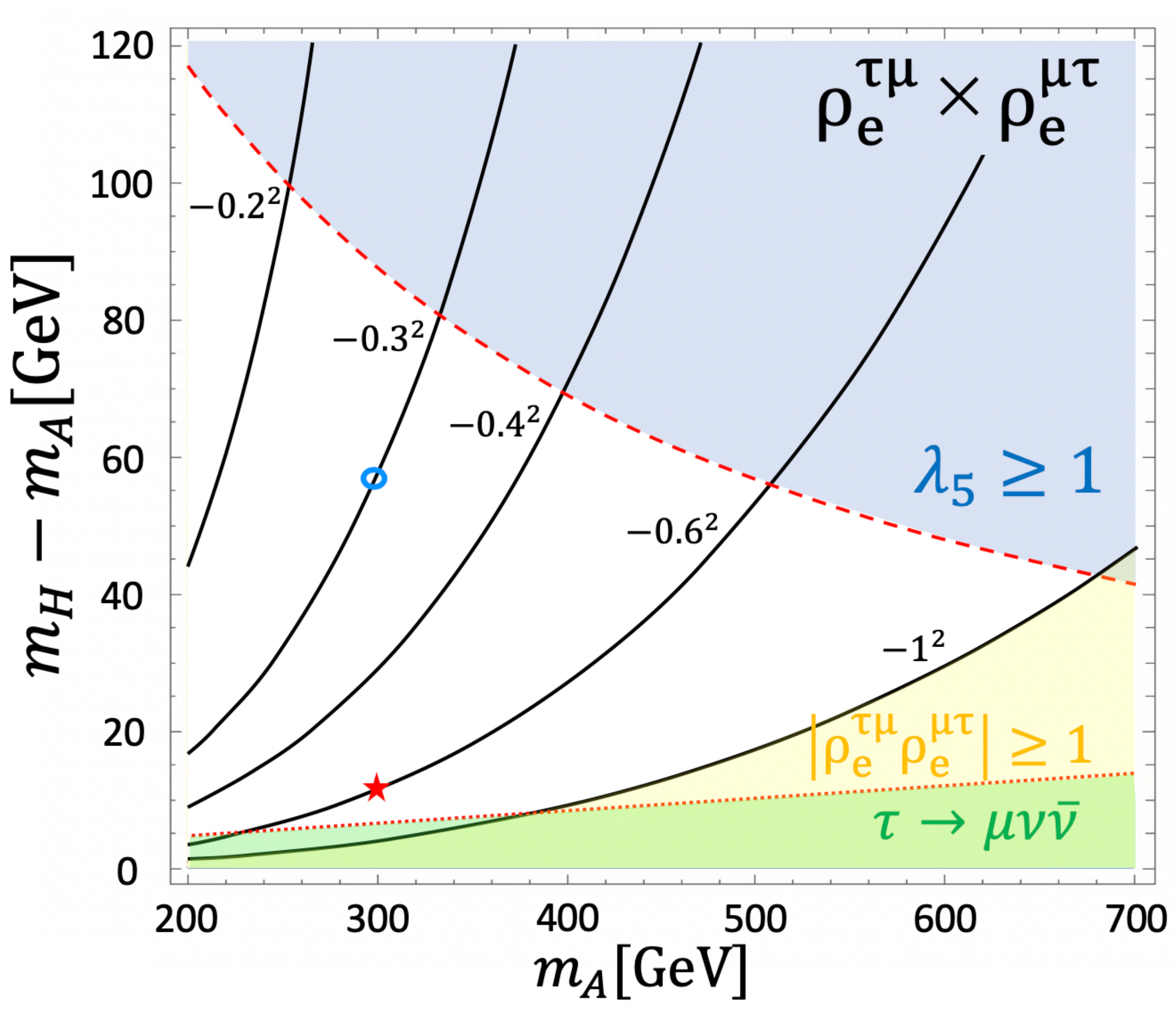}
            \caption{ The required value of $\rho_{e}^{\mu\tau} \times \rho_{e}^{\tau\mu}$ to obtain $\delta a_\mu=2.8\times 10^{-9}$. The gray shaded region corresponds to $\lambda_5\ge 1$. The green shaded region is excluded by the $\tau\to\mu\nu\bar\nu$ process mediated by the charged Higgs where we assumed $m_{H^\pm}=m_H$. The yellow shaded region corresponds to $|\rho_{e}^{\mu\tau}\rho_{e}^{\tau\mu}| \ge 1$. 
 The benchmark points adopted for the LHC study in the following section are indicated by the blue circle (BP1) and the red star (BP2). }
    \label{rhoNmassdif}
  \end{center}
\end{figure}

In the following, we consider the mass spectrum of the scalars that  satisfies $m_A\le m_H=m_{H^\pm}$.
Based on Refs.~\cite{Omura:2015nja,Omura:2015xcg}, we assume that those masses are in a 
few hundreds GeV.
Fig.~\ref{rhoNmassdif} shows the required value of the product $\rho_{e}^{\mu\tau}\rho_{e}^{\tau\mu}$ 
to obtain $\delta a_\mu=2.8\times10^{-9}$ in the ($m_A$, $\Delta_{H-A}$)-plane. The gray shaded region corresponds
to $\lambda_5\ge1$. The charged Higgs contributes to the $\tau\to\mu\nu\nu$ 
process~\cite{Omura:2015xcg}, and the corresponding region excluded by its measurements is 
shown in the green region. The yellow shaded region indicates $|\rho_e^{\tau\mu}\rho_e^{\mu\tau}| \ge1$.
From the plot, we see  $\rho_{e}^{\mu\tau}\rho_{e}^{\tau\mu}$ is required to be ${\cal O}(1)$ in order to obtain 
a sufficient contribution to explain the deviation in $ a_\mu$
\footnote{In the type-II 2HDM, that is widely discussed, $\rho_e^{\tau \tau}$ is, for instance, given by
$\rho_e^{\tau \tau}=\frac {\sqrt 2 m_\tau}{v} \tan \beta$. $\tan \beta$ corresponds to the ratio of two Higgs VEVs. If $\rho_e^{\mu\tau}$
is normalized by $\frac {\sqrt 2 m_\tau}{v} $ like $\rho_e^{\mu\tau}=\frac {\sqrt 2 m_\tau}{v} {\rm{tan}}\beta$, $\tan \beta$ is estimated as about one hundred when $\rho_e^{\mu\tau}=1$. }. 
We also see that $10$~GeV $\lesssim \Delta_{H-A} \lesssim 100$~GeV 
is required. The allowed region of $\Delta_{H-A}$ shrinks as $m_A$ gets larger.  When we further require $|\rho_{e}^{\mu\tau}\rho_{e}^{\tau\mu}| \le 1$, the allowed region is limited as $m_A \lesssim 680$~GeV.
%%%%%%%%%%%%%%%%%%%
%%%%%%%%%%%%%%%%%%%%%

%%%%%%%%%%%%%%%%%%%%%%%%%%%%%%%%%%%%
%%%%%%%%%%%%%%%%%%%%%%%%%%%%%%%%%%%%
\section{Collider Signals at the LHC}
\label{sec:lhc}
\subsection{multi-lepton signatures}
\begin{figure}[t]
  \begin{center}
    \hfill
    \includegraphics[width=5cm]{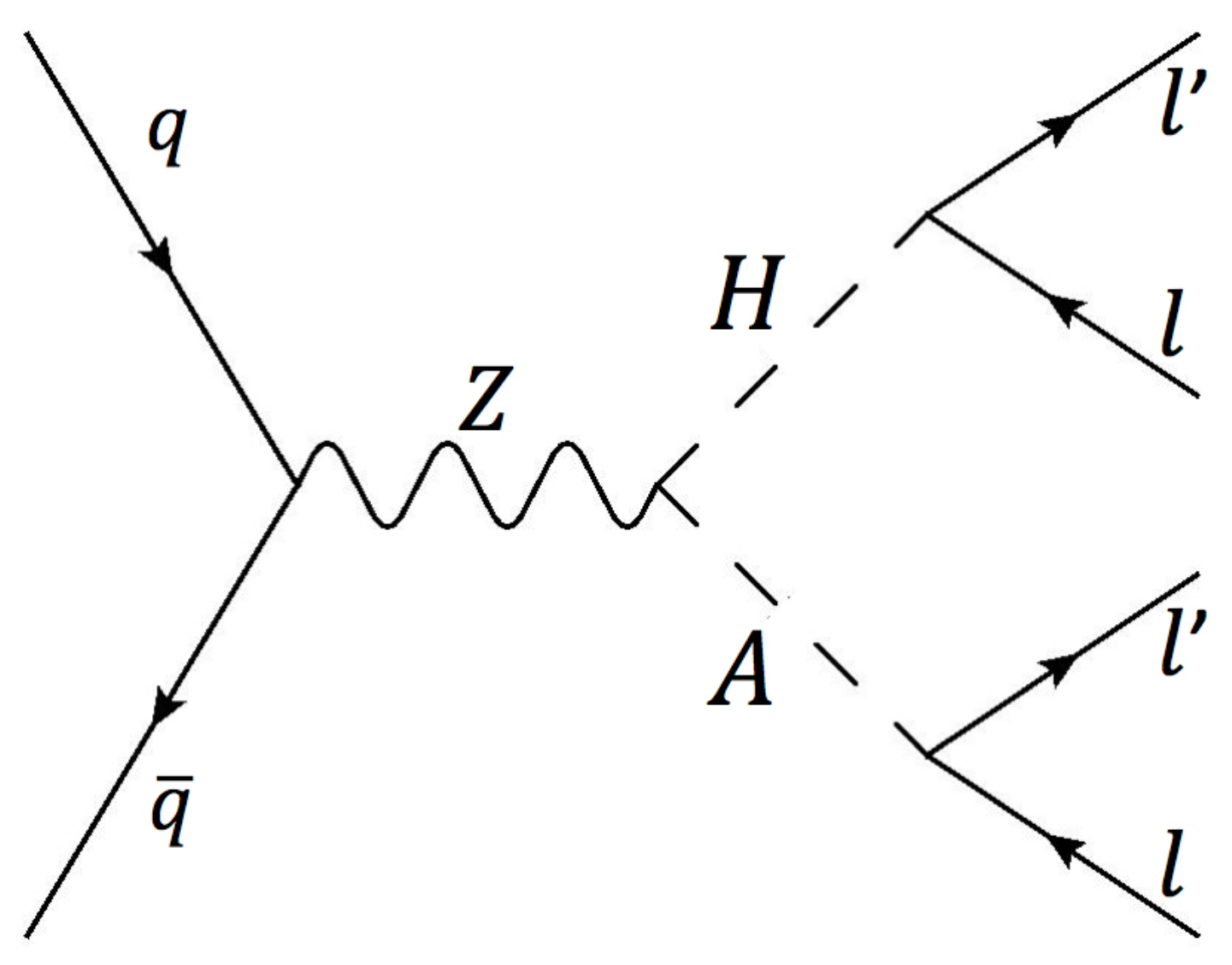}
    \hfill
    \includegraphics[width=5cm]{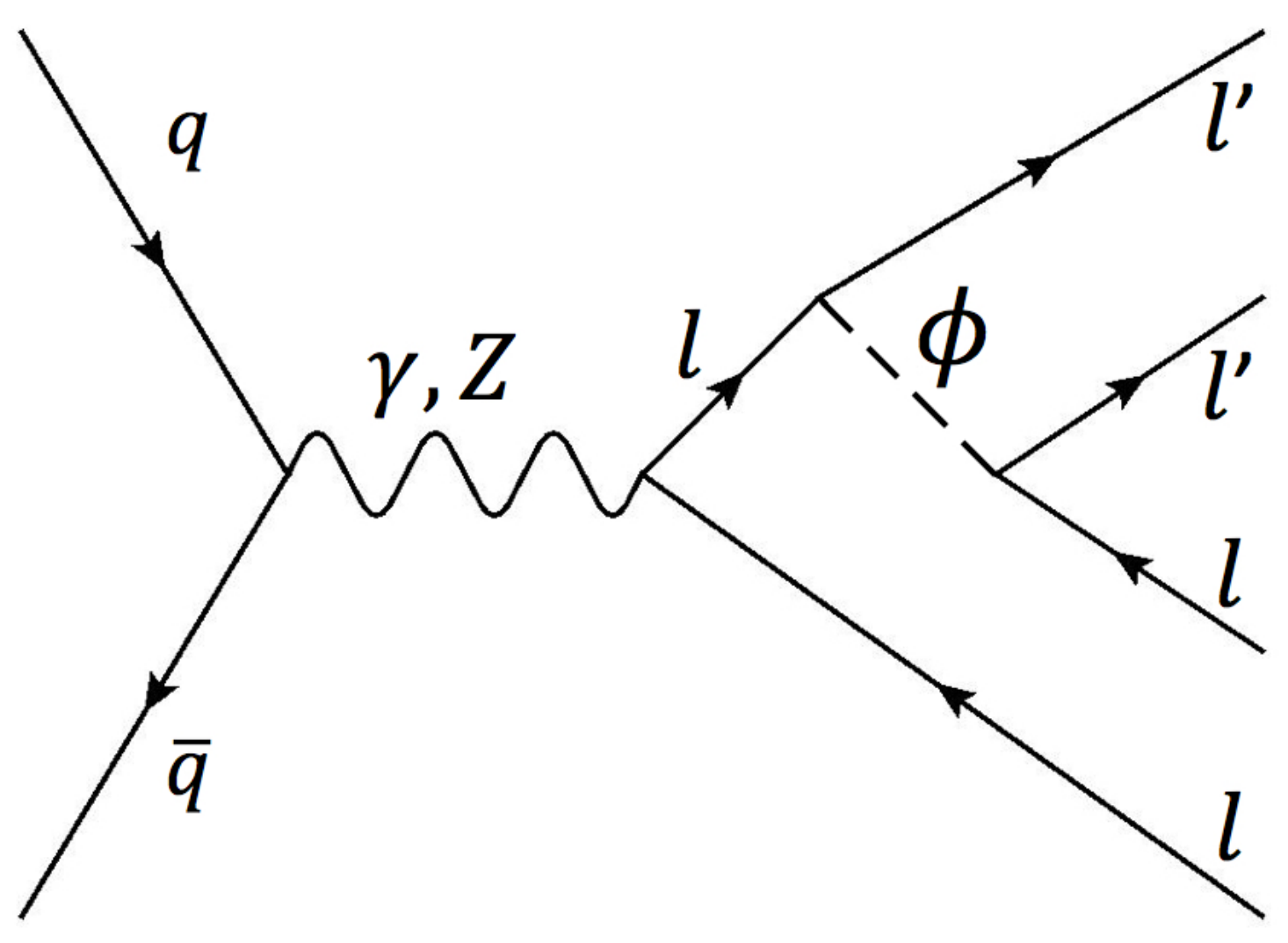}
    \hfill
    \ 
    \caption{ The representative diagrams contributing to the 4 lepton channel in our model.}
    \label{diagrams}
          \end{center}
\end{figure}

As we discussed in Sec.~\ref{muong2}, only $\rho_e^{\mu\tau}$ and $\rho_e^{\tau\mu}$ are the relevant parameters in the Yukawa matrix
in our scenario.
With these minimal entries in the Yukawa matrix, it is relatively difficult to search for the additional Higgs 
bosons at the LHC
as they do not couple to the valence quarks.
Even without any valence quark coupling, additional scalars originated from the two Higgs doublets 
can be produced in pair at the LHC via the Drell-Yan processes induced 
by the electroweak interaction. 
Each extra Higgs boson decays into leptons in a flavor-violating way, and therefore, they provide the multi-lepton final states 
as depicted in the left diagram in the Fig.~\ref{diagrams}.
In principle, the right diagram in Fig.~\ref{diagrams} also contributes to the 4 lepton channel, however, 
it is negligible for the region of our interest.~\footnote{We have explicitly checked it by 
varying $|\rho_e^{\mu\tau}|=|\rho_e^{\tau\mu}|$ from 0.01 to 1 for $m_A=m_H=300$ GeV.}
Already the ATLAS and the CMS have collected the LHC run 2 data of about $150$~fb$^{-1}$, which makes the exotic searches 
with such a low cross section but enjoying the low SM background (SMBG) very promising.

We consider the three production processes, $HA$, $\phi H^\pm$, and $H^+H^-$, where $\phi=H, A$. In our setup, both the neutral Higgs $H$ and $A$ decay into $\tau^\pm\mu^\mp$, while $H^\pm$ decays into $\tau^\pm \nu$ and  $\mu^\pm \nu$. Therefore, those processes will end up as multi-lepton and multi-tau final states.
Especially, the novel final state, the same-sign two muons and the same sign two taus: 
$\mu^\pm \mu^\pm\tau^\mp\tau^\mp$, would be the very characteristic signature with essentially no SMBG. 
 Note that the lepton flavor violating LHC signatures in our scenario is different from the ones in the lepton-specific 2HDM~\cite{Chun:2015hsa}, where the scalar $A$ is very light to explain the $a_\mu$ anomaly. Therefore, the signatures induced by the extra scalar involve $A \to \tau\tau$, so that they are the multi-tau signatures.

%%%%%%
%%%%%%

\subsection{Current constraints}
\begin{figure}[h!t]
  \begin{center}
    \includegraphics[width=10cm]{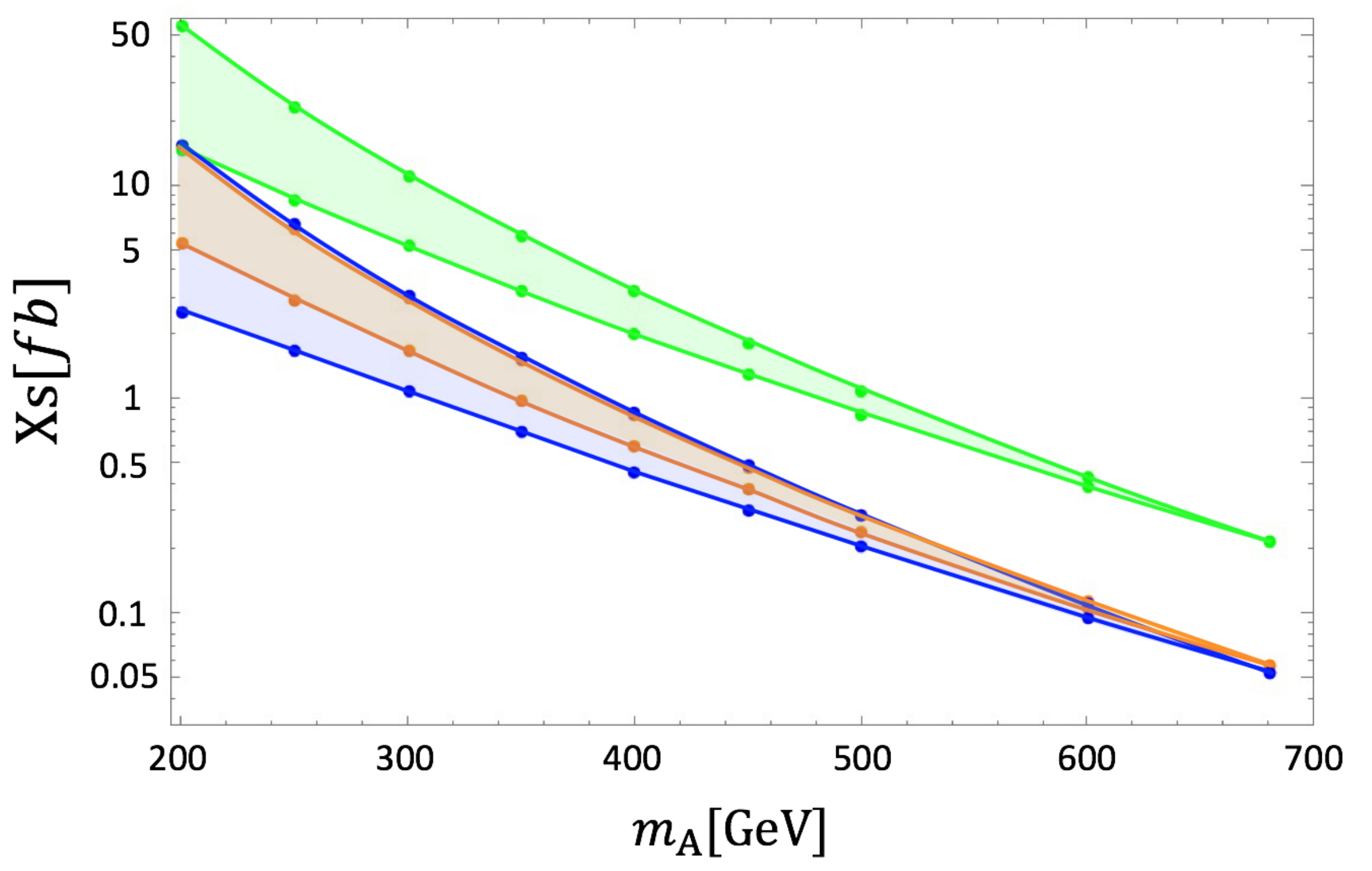}
    \caption{The pair-production cross sections for $\phi H^{\pm}$ ($AH^\pm$ and $HH^\pm$ are summed, green hatched),
    $H^-H^+$ (blue hatched), and $HA$ (orange hatched) at the LHC~with $\sqrt s=13$ TeV as functions of $m_A$. 
    In the each process, the upper line is given by assuming $|\rho_e^{\mu\tau}|=|\rho_e^{\tau\mu}|=1$, that corresponds to the minimum $\Delta_{H-A}$, while the lower line is obtained by $\lambda_5=1$.
             }
    \label{Xs_result0}
  \end{center}
\end{figure}

Fig.~\ref{Xs_result0} shows the pair production cross sections for the three processes, $\sigma(\phi H^\pm)$, $\sigma(HA)$, and $\sigma(H^+ H^-)$ at the LHC $13$~TeV as a function of $m_A$ in the green band, in the orange band, and in the blue band, respectively.
Although there are the five parameters, $\rho_e^{\mu\tau},\rho_e^{\tau\mu}, m_A, m_H$, and $m_{H^\pm}$, in our setup,
the cross sections depend only on the relevant masses but not on either $\rho_e^{\mu\tau}$ or $\rho_e^{\tau\mu}$  
since the scalars are produced via the weak interaction.
Following the discussion in the previous section, $\Delta_{H-A}$ for each $m_A$ value is constrained by the perturbativity of the parameters.
%,
%which is why each cross section is shown as the band. 
Then, we plot our prediction based on the allowed region in Fig. \ref{rhoNmassdif}.
The upper line corresponds to the possible minimum value for $m_H (=m_{H^{\pm}})$, that comes from $|\rho_{e}^{\mu\tau}|,|\rho_{e}^{\tau\mu}| \le 1$, 
and the lower line corresponds to the maximum related with the $\lambda_5\le 1$ constraint.

\begin{figure}[h]
  \begin{center}
    \includegraphics[width=10cm]{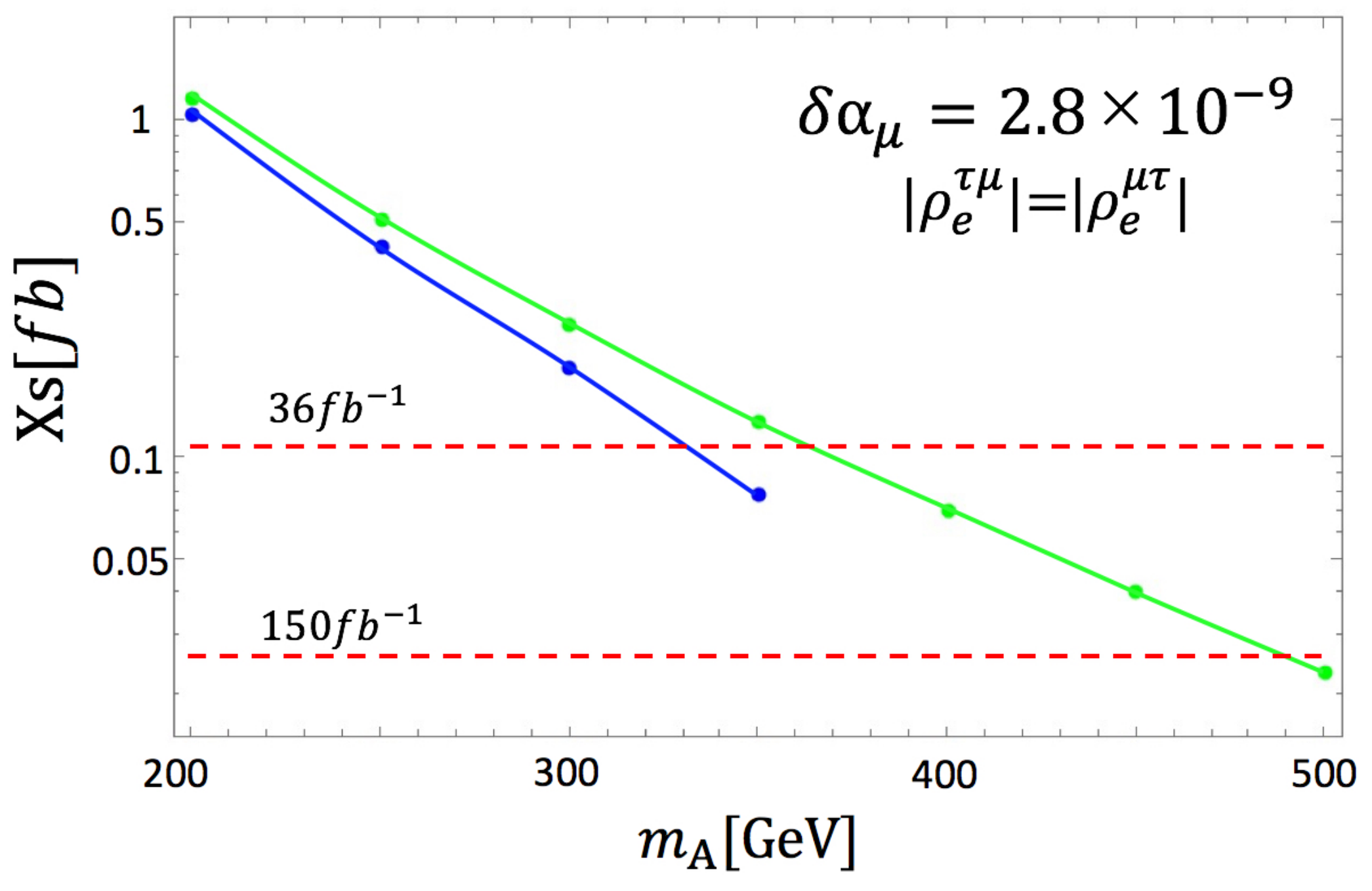}
            \caption{ The $\tau^\pm\tau^\pm \mu^\mp\mu^\mp$ signal cross section after the selection cut as a function of $m_A$. 
            We show the two cases $|\rho_e^{\tau\mu}|=|\rho_e^{\mu\tau}|=0.3$~(blue), and $0.6$~(green). 
            The other parameters are fixed to reproduce the required deviation of 
            $\delta a_\mu = 2.8 \times 10^{-9}$. }
    \label{Xs_result}
  \end{center}
\end{figure}

For the multi-lepton signatures, we have to consider the branching ratios, 
which are
controlled by the $\rho_e^{\mu\tau}$ and $\rho_e^{\tau\mu}$ for $H^\pm$
while independent for $\phi$ ($= A$ and $H$) as follows,
\begin{eqnarray}
&&BR(\phi \to \tau^+ \mu^-) = BR(\phi \to \tau^- \mu^+) = 0.5, \cr
&&BR(H^\pm \to \tau^\pm \nu)  =   1 - BR(H^\pm \to \mu^\pm \nu)  =
\frac{|\rho^{\mu\tau}_e|^2}{|\rho^{\tau\mu}_e|^2+|\rho^{\mu\tau}_e|^2} \equiv r.
\label{br}
\end{eqnarray}
Depending on the branching ratios, the resulting fraction of the multi-lepton final states is determined.
Fig.~\ref{Xs_result} shows the rough estimate of the expected number of the $\mu^\pm\mu^\pm\tau^\mp\tau^\mp$ signal events at the LHC~$13$~TeV as a function of $m_A$, where we 
consider only the contribution from $HA$ production, and $m_H$ is taken to obtain $\delta a_\mu=2.8\times10^{-9}$
for the two cases with $|\rho^{\mu\tau}_e|=|\rho^{\tau\mu}_e|=0.6$ and $0.3$.

We generate the signal events using {\tt MadGraph5}~\cite{Alwall:2014hca}
to estimate the effect of the minimal acceptance cut, $|\mathbf{p}_{T,\mu}|,|\mathbf{p}_{T,\tau}|\ge 20$~GeV, $|\eta_{\mu}|,|\eta_{\tau} |\le 2.7$, and $\Delta R \ge 0.1$ for all pair of charged leptons.
We assume the hadronic tau-tagging efficiency of $70\%$ and an excellent tau charge reconstruction~\cite{tautag1}. 
For the mass scale we consider, taus from $H$ and $A$ decays are expected to be highly boosted,
and therefore, the constituents of the tau-jet are highly collimated~\cite{Rainwater:1998kj}. It makes the tau easier to capture experimentally.
Taking the hadronic tau decay branching ratio of about 65\% into account, roughly  50\% of a tau would be tagged as a tau-jet.

We expect the discrimination power of the signal against the SMBG in the $\mu^\pm\mu^\pm \tau^\mp\tau^\mp$ mode is much 
better than the one in the Ref. \cite{Altmannshofer:2016brv},  where the $\mu^\pm\mu^\pm \tau^\mp\tau^\mp$ signal that only one of $\mu^\pm \tau^\mp$ comes from a heavy resonance is considered. 
Hence, we especially assume the SMBG in the  $\mu^\pm\mu^\pm \tau^\mp\tau^\mp$ mode is negligible. 
We estimate the signal significance by $\sqrt{\sigma {\cal L}}$, where $\sigma$ and ${\cal L}$ denote the signal cross section after the selection cut and the integrated luminosity, respectively. The red-dashed lines in Fig.~\ref{Xs_result} represent the cross sections corresponding to the significance $2 \sigma$ for 36~fb$^{-1}$ and 150~fb$^{-1}$, corresponding to 0.11 fb and 0.027 fb, respectively. 
Therefore, the current LHC data would be enough sensitive to the $m_A \sim 500$~GeV.
We note that the $\mu^\pm\mu^\pm\tau^\mp\tau^\mp$ signatures are predicted also by other models~\cite{Chiang:2017vcl,Abe:2019bkf}. 
Slepton searches constrain the charged Higgs mass as their quantum charges are identical. 
The latest stau searches at the LHC with the $139$~fb$^{-1}$ data in the $2\tau + E\!\!\!/_T$ mode 
excludes the stau mass between $150$~GeV and $300$~GeV for
 $BR({\tilde \tau} \to \tau \tilde\chi)=100~\%$~\cite{ATLAS:2019ucg}.  The lower bound on slepton mass is already 
 around $700$~GeV using the same integrated luminosity, but it 
assumes the degenerate four sleptons ${\tilde l}={\tilde e}_L, ~{\tilde e}_R, ~{\tilde \mu}_L$ and ${\tilde \mu}_R
$ and $BR({\tilde l} \to l \tilde\chi)=100~\%$~\cite{ATLAS:2019cfv}, therefore, not applicable to our case 
directly. Although the results for $36$~fb$^{-1}$ is currently only available, the CMS provides the lower bound on the left-handed 
smuon mass to be $280$~GeV assuming $BR({\tilde \mu}_L \to \mu \tilde\chi)=100~\%$~\cite{Sirunyan:2018nwe}.
Although these results would constrain our model in principle, there is no explicit study yet for 
the case with the intermediate branching ratio, which is relevant to our setup especially for 
$|\rho_e^{\tau\mu}| \simeq |\rho_e^{\mu\tau}|$. In that case,  searches for the $\tau+\mu$ plus 
missing momentum signatures would be desired.

%%%%%%%%%%%%%%%%%%%%%%%%%%%%%%%%%%%%
%%%%%%%%%%%%%%%%%%%%%%%%%%%%%%%%%%%%
%%%%%%%%%%%%%%%%%%%%%%%%%%%%%%%%%%%%
\subsection{Future Prospects}
%%%%%%%%%%%%%%%%%%%%%%%%%%%%%%%%%%%%
%%%%%%%%%%%%%%%%%%%%%%%%%%%%%%%%%%%%
%%%%%%%%%%%%%%%%%%%%%%%%%%%%%%%%%%%%
Once the LHC accumulates enough data, the mass reconstruction of the extra Higgses would be possible.
For the illustration purpose, we select the two benchmark points and show how to reconstruct 
the mass spectrum in this scenario. The values of the mass parameters and the relevant cross 
sections at the LHC at $\sqrt s=14$ TeV are summarized in Tab.~\ref{bp}. We generate the signal 
events at the LHC assuming $\sqrt s=14$~TeV using {\tt MadGraph5}~\cite{Alwall:2014hca} and 
{\tt PYTHIA8}~\cite{Sjostrand:2006za}. Then, the events are interfaced 
to {\tt DELPHES3}~\cite{deFavereau:2013fsa} for the fast detector simulation. We consider the 
three categories of the signal processes $HA$, $\phi H^\pm$, and $H^+H^-$, and we expect 
that they are the main contributions for the 4 lepton, 3 lepton, and 2 lepton events. 
Note that tau is included in leptons in our definition. 
As an acceptance cut, we require,  $|\mathbf{p}_{T,\mu}|,|\mathbf{p}_{T,\tau}|> 20$~GeV, $ |\mathbf{p}_{T,j}|> 30$~GeV, 
and $|\eta_{e,\mu,j}| < 2.4$.

\begin{table}[t]
\begin{tabular}{l|rrr| rrrrr}
\hline
        & $m_A$ & $m_H$ & $m_{H^{\pm}}$ & $\sigma(HA)$ & $\sigma(AH^\pm)$ & $\sigma(HH^\pm)$& $\sigma(H^+H^-)$  \cr
\hline
BP1 & 300 GeV& 358 GeV& 358 GeV& 2.4 fb& 4.6 fb& 3.3 fb& 1.8 fb\cr
BP2 & 300 GeV& 312 GeV& 312 GeV& 3.3 fb& 6.3 fb& 5.7 fb& 3.2 fb \cr
\hline
\end{tabular}
\caption{
The mass parameters of the two benchmark points and the production cross sections at LHC~14~TeV.
For $\sigma(AH^\pm)$ and $\sigma(HH^\pm)$,
each $H^+$ contribution is roughly the twice the corresponding $H^-$ contribution due to the PDF effects,
and both contributions are summed.
}
\label{bp}
\end{table}

%%%%%%%%%%%%%%%%%%%%%%%%%%%%%%%%%%%%
%%%%%%%%%%%%%%%%%%%%%%%%%%%%%%%%%%%%
\subsubsection{4 lepton modes}
%%%%%%%%%%%%%%%%%%%%%%%%%%%%%%%%%%%%
%%%%%%%%%%%%%%%%%%%%%%%%%%%%%%%%%%%%
First let us consider the 4 lepton final states from the $HA$ production. 
Each $A$ and $H$ decays into $\tau^+ \mu^-$ and  $\tau^- \mu^+$ at $50\%$ each, 
thus the half provides the same-sign di-$\mu$ and the same-sign di-$\tau$ events ($\mu^\pm\mu^\pm \tau^\mp\tau^\mp$), 
and the other half provides the opposite-sign di-$\mu$ and the opposite-sign di-$\tau$ events ($\mu^+\mu^- \tau^+\tau^-$). 
After applying the acceptance cut selecting two isolated muons and two tau-tagged jets, about $9~\%$ of the events pass the acceptance cut.
We name them, $\mu_1, \mu_2, \tau_1^{\rm vis}$, and $\tau_2^{\rm vis}$ in $p_T$-order.

To reconstruct the two $\tau \mu$ resonances, in the former case we have to consider two possible combinations, while no such a problem arises in the latter case. Although we can use just the both combinations to identify the peaks in the $\mu^\pm\mu^\pm \tau^\mp\tau^\mp$ events 
as the contribution from the wrong combinations just provides the continuum distributions, 
to obtain the clear peaks to estimate the mass resolution, 
we further drop the one combination event-by-event basis using the $\chi^2$ value defined as follows.

\begin{figure}[t]
  \begin{center}
    \includegraphics[width=5cm]{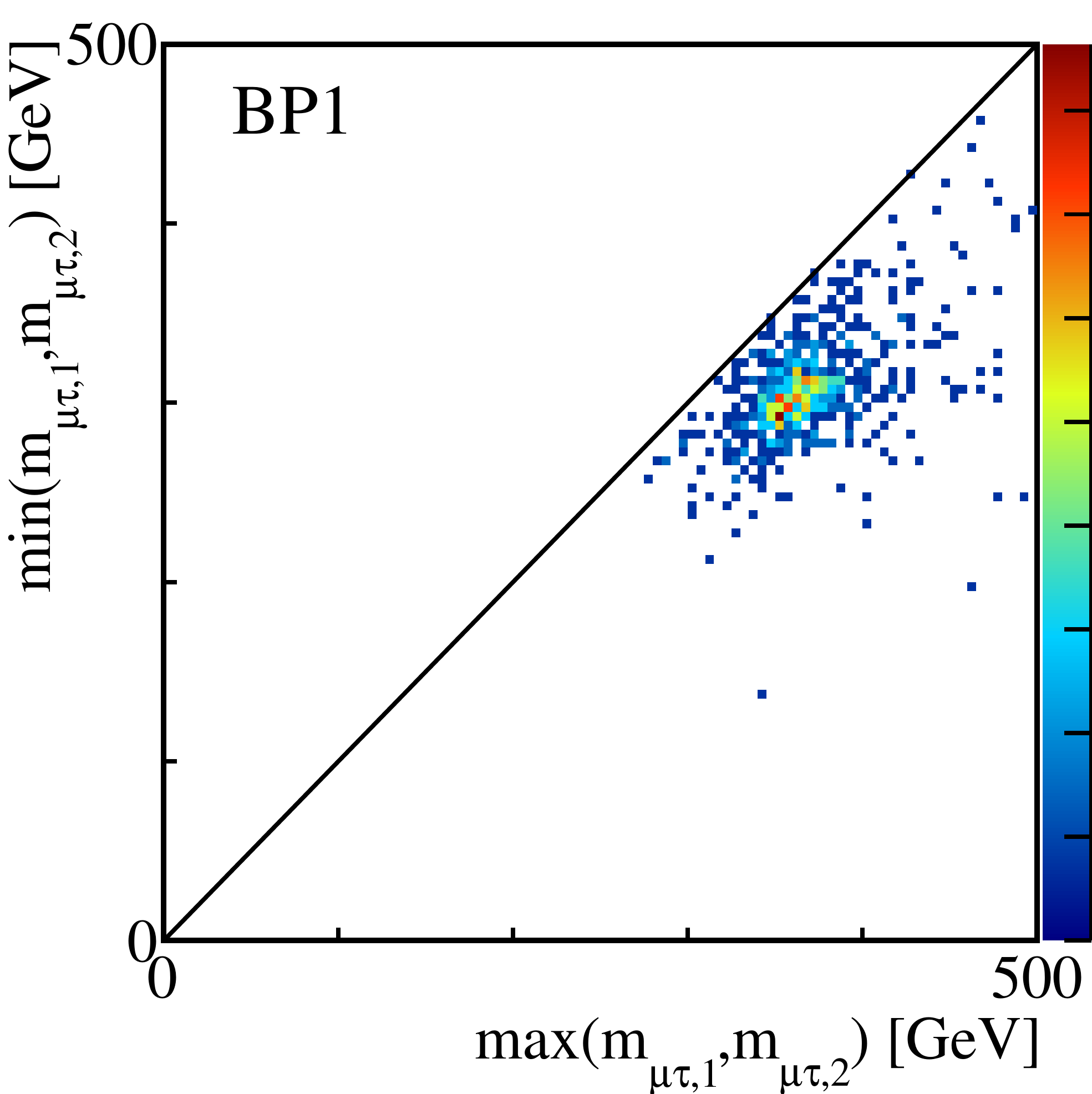}
    \includegraphics[width=5cm]{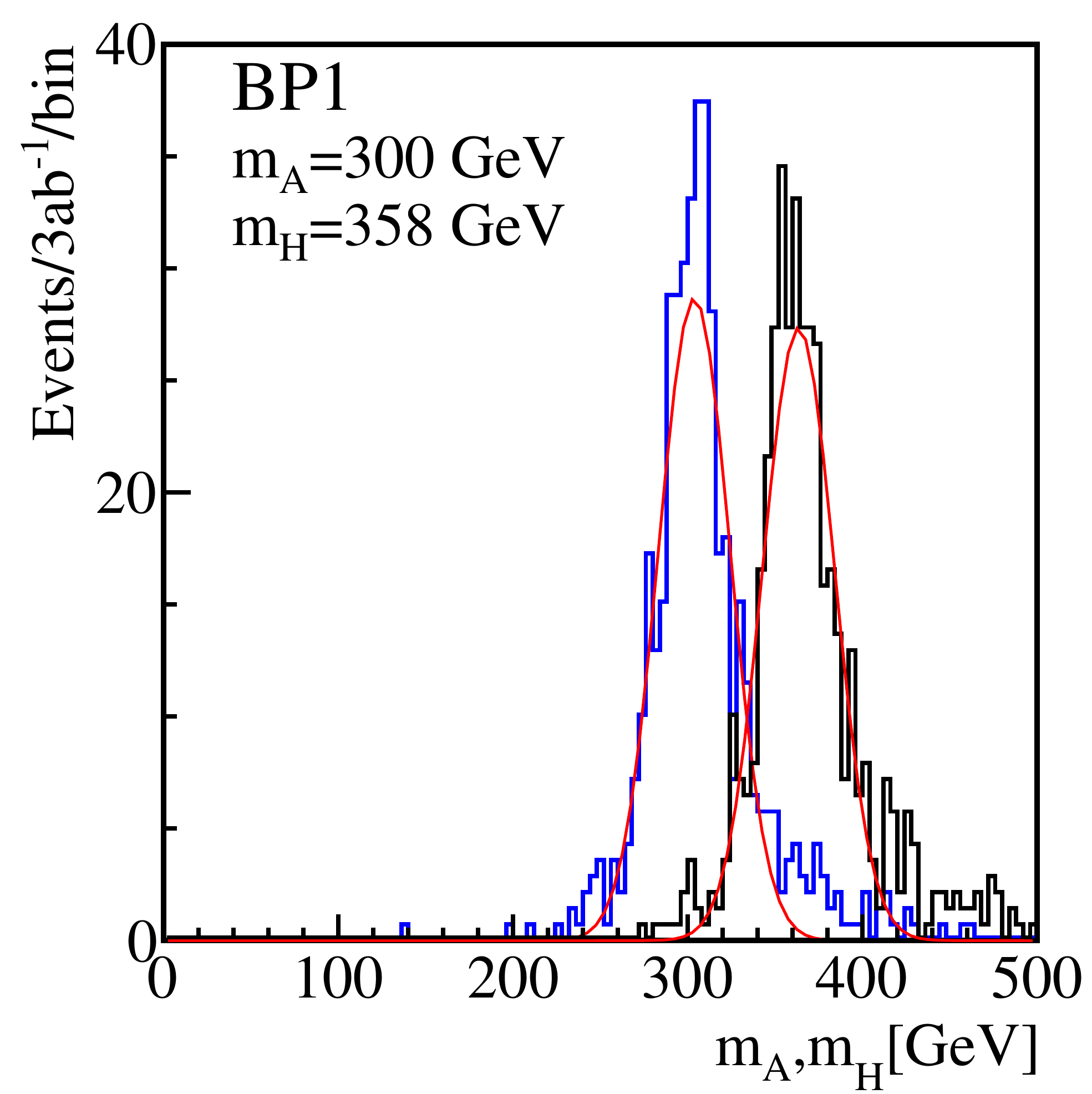}
    \includegraphics[width=5cm]{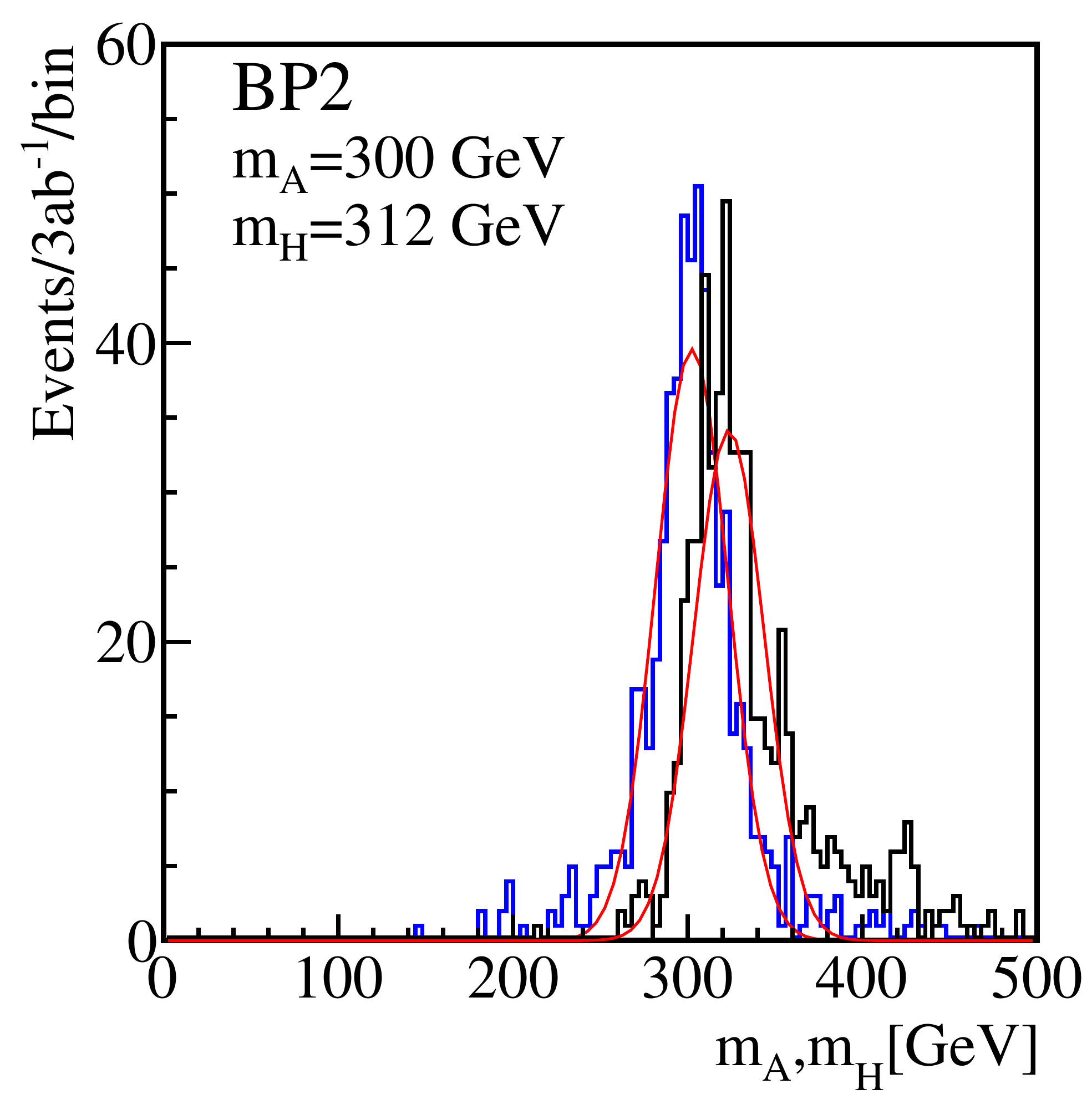}
            \caption{ The 2-dimensional $m_{\mu\tau}^{\min}$ vs. $m_{\mu\tau}^{\max}$ distribution 
            from $HA$ production (left). Denser regions are depicted in red points. Reconstructed 
            $m_A$ and $m_H$ distributions for BP1 (center) and BP2 (right).}
    \label{4lep}
  \end{center}
\end{figure}

As a visible hadronic tau-jet carries only a part of the original tau momentum due to the escaping neutrino momentum, 
we adopt the collinear approximation~\cite{Rainwater:1998kj}  to reconstruct the original tau momenta with the help of the transverse missing momentum, which are $\mathbf{p}_{\tau_i} = (1 + c_i)  \mathbf{p}^{\rm vis}_{\tau_i}$ for $i=1,2$ satisfying
\begin{eqnarray}
\mathbf{p}\!\!\!/_T = c_1 \mathbf{p}^{\rm vis}_{T, \tau_1} + c_2  \mathbf{p}^{\rm vis}_{T,\tau_2}\ \ \ \  (c_1, c_2> 0).
\label{coll}
\end{eqnarray}
The idea is that the momentum carried by the neutrino is aligned to the visible momentum, which is 
better when the original $\tau$ is boosted.
Here, we require $E\!\!\!/_T =|\mathbf{p}\!\!\!/_T |>10$~GeV and only accept events where Eq.(\ref{coll}) has a solution, which further loses 30~\% of events.
We reconstruct the two invariant masses in the two possible combinations: 
\begin{eqnarray}
{\rm combination\ } 1 &:& m_{\mu_1\tau_1} \ \ {\rm and} \ \ m_{\mu_2\tau_2} \cr
{\rm combination\ } 2 &:& m_{\mu_1\tau_2} \ \ {\rm and} \ \ m_{\mu_2\tau_1} 
\end{eqnarray}
For each combination $i$, we name the smaller one as $m_{\mu\tau,i}^{\min}$ and 
the larger one $m_{\mu\tau,i}^{\max}$. We define the hypothetical $\chi^2_{i}(m_A, m_H)$ as 
\begin{eqnarray}
\chi^2_{i}(m_A, m_H)=(m_{\mu\tau,i}^{\min}- m_A)^2/\sigma_{\rm res}^2 + 
(m_{\mu\tau,i}^{\max} - m_H)^2/\sigma_{\rm res}^2,
\end{eqnarray}
and select the combination event-by-event which minimizes 
the sum of $\min(\chi^2_{1},\chi^2_{2})$.
The 2-dimensional distribution in the $m_{\mu\tau}^{\min}$ vs. $m_{\mu\tau}^{\max}$ plane after selecting the one combination minimizing the sum of the $\chi^2$ 
is shown in the left panel of Fig.~\ref{4lep}.
Note that the denser regions are depicted in red points.
The projected distributions along  $m_{\mu\tau}^{\min}$ and $m_{\mu\tau}^{\max}$ axes, which supposedly 
corresponds to the reconstructed $m_A$ and $m_H$ distributions, are shown for the benchmark point 1 (BP1) 
in the central panel, and for the benchmark point 2 (BP2) in the right panel. 
After the acceptance cut, ${\cal O}(250)$ events for BP1 (${\cal O}(300)$ events for BP2) 
remains for 3~ab$^{-1}$.

Based on our simulation, the peak is smeared due to the incomplete tau momentum reconstruction but 
still the mass reconstruction resolution $\sigma_{\rm res}$ is about 20~GeV, therefore, 
$\Delta_{H-A} = 58$~GeV in BP1 would be easily separated, where the fitted reconstructed 
mass difference is 60~GeV. We also show the mass separation for the BP2 with 
$\Delta_{H-A}=12$~GeV on the right panel in Fig.~\ref{4lep}, where the fitted reconstructed 
mass difference is 20~GeV. 
It shows that the algorithm tends to separate the two peaks if the mass difference is smaller 
than the intrinsic resolution. Nevertheless, since most of the relevant parameter space provides 
an enough mass difference as shown in Fig.~\ref{rhoNmassdif}, it would not be a problem in the 
most region.
%%%%%%%%%
%%%%%%%%%

%%%%%%%%%%%%%%%%%%%%%%%%%%%%%%%%%%%%
%%%%%%%%%%%%%%%%%%%%%%%%%%%%%%%%%%%%
\subsubsection{3 lepton modes}
%%%%%%%%%%%%%%%%%%%%%%%%%%%%%%%%%%%%
%%%%%%%%%%%%%%%%%%%%%%%%%%%%%%%%%%%%
Next, we consider the 3 lepton final states, which are mainly produced by $\phi H^\pm$ processes, 
where $\phi$ (= $A$ and $H$) decays into $\tau^\pm\mu^\mp$, and $H^\pm$ decays into $\tau^\pm \nu$ 
or $\mu^\pm\nu$, whose ratio is controlled by the $\rho^{\mu\tau}_e/\rho^{\tau\mu}_e$ as in Eq.(\ref{br}). 
Therefore, through the 3 lepton events, we can access the information on the ratio 
$\rho^{\mu\tau}_e/\rho^{\tau\mu}_e$ by measuring the ratio of $2\mu1\tau$ and $1\mu2\tau$ 
events as well as the information on $m_{H^\pm}$. 
In this mode, the complication comes from three reasons. 
First, two possible $\tau\mu$ resonances $A$ and $H$ 
with different masses can contribute to the same event topology. 
Second, due to the neutrino contribution, the $H^\pm$ mass is 
not able to be reconstructed using the invariant mass. 
Third, the intrinsic $2\tau1\mu$ events contribute to $1\tau2\mu$
events due to the $\tau \to \mu\nu\nu$ decay. 
%%%%%%%%%
\begin{figure}[t]
  \begin{center}
    \includegraphics[width=5cm]{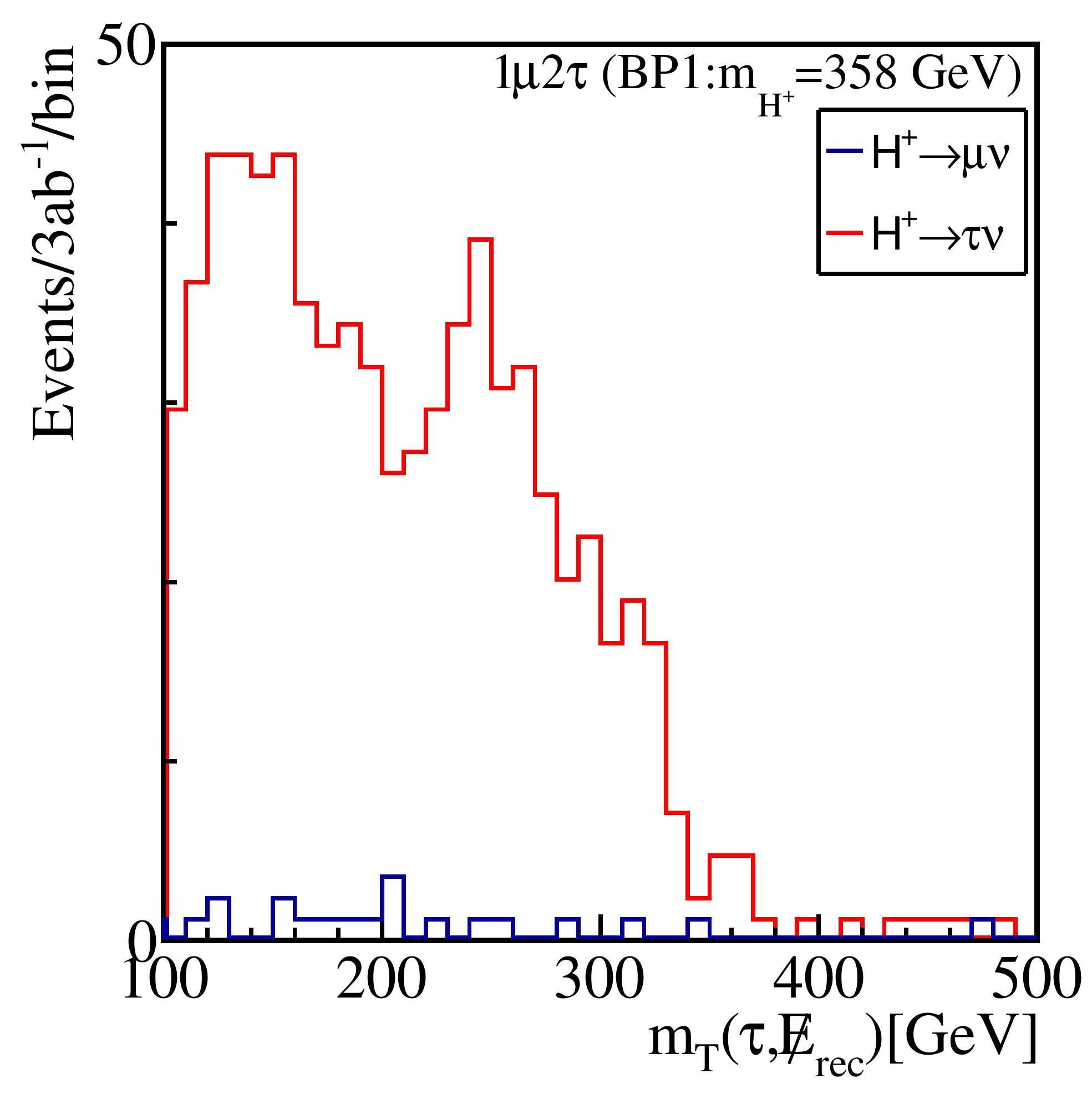}
    \includegraphics[width=5cm]{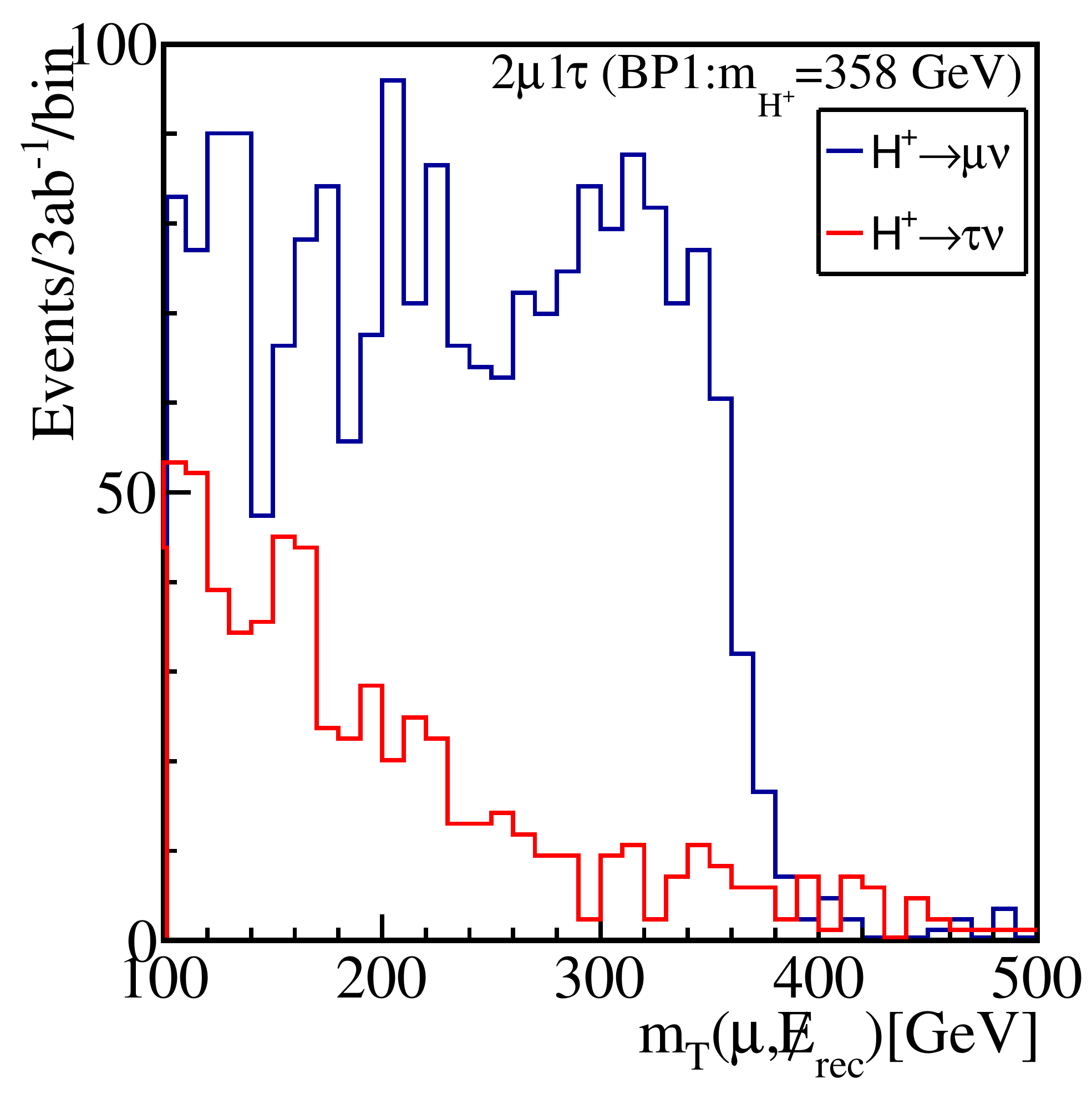}
    \includegraphics[width=5cm]{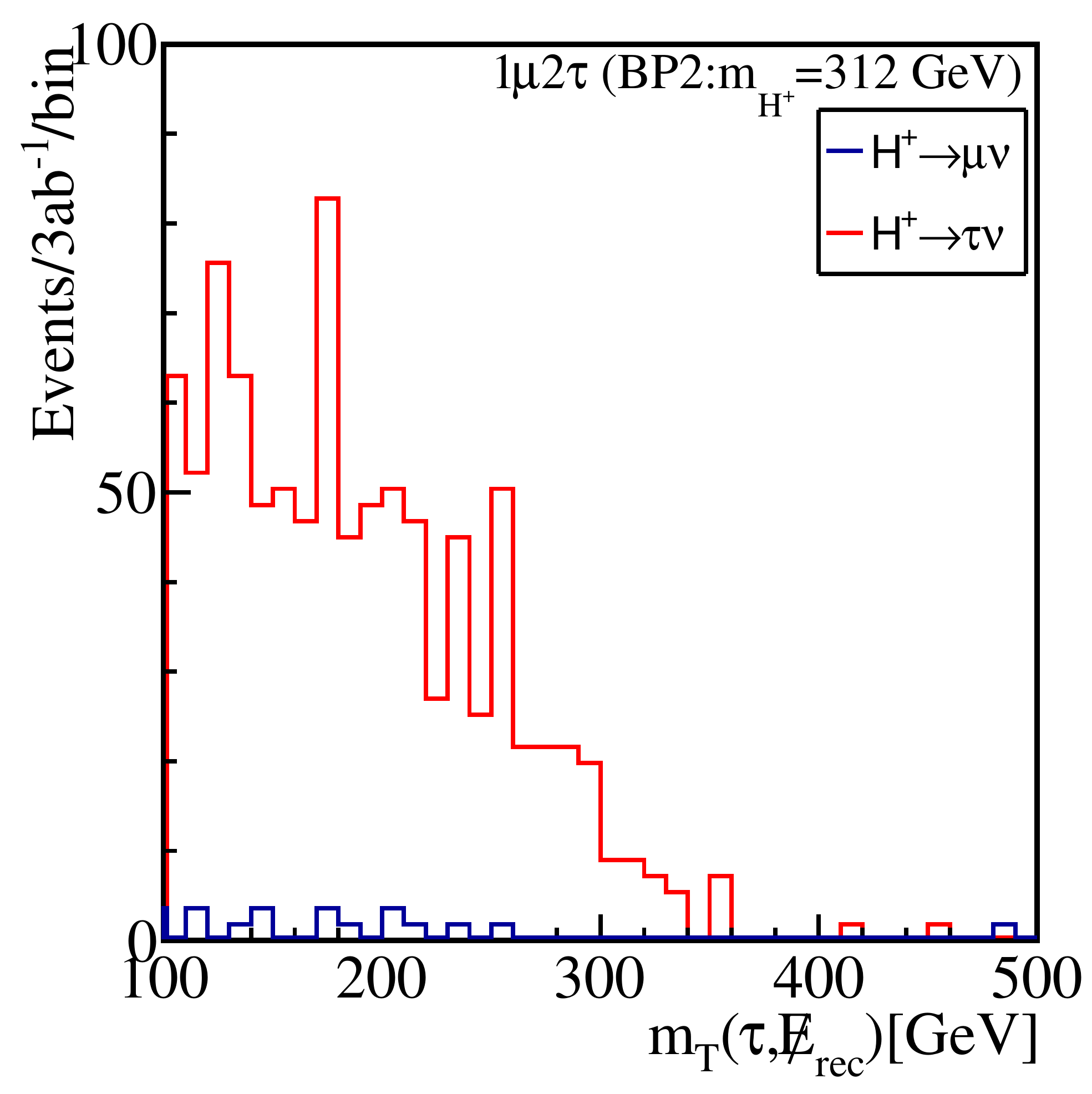}
    \includegraphics[width=5cm]{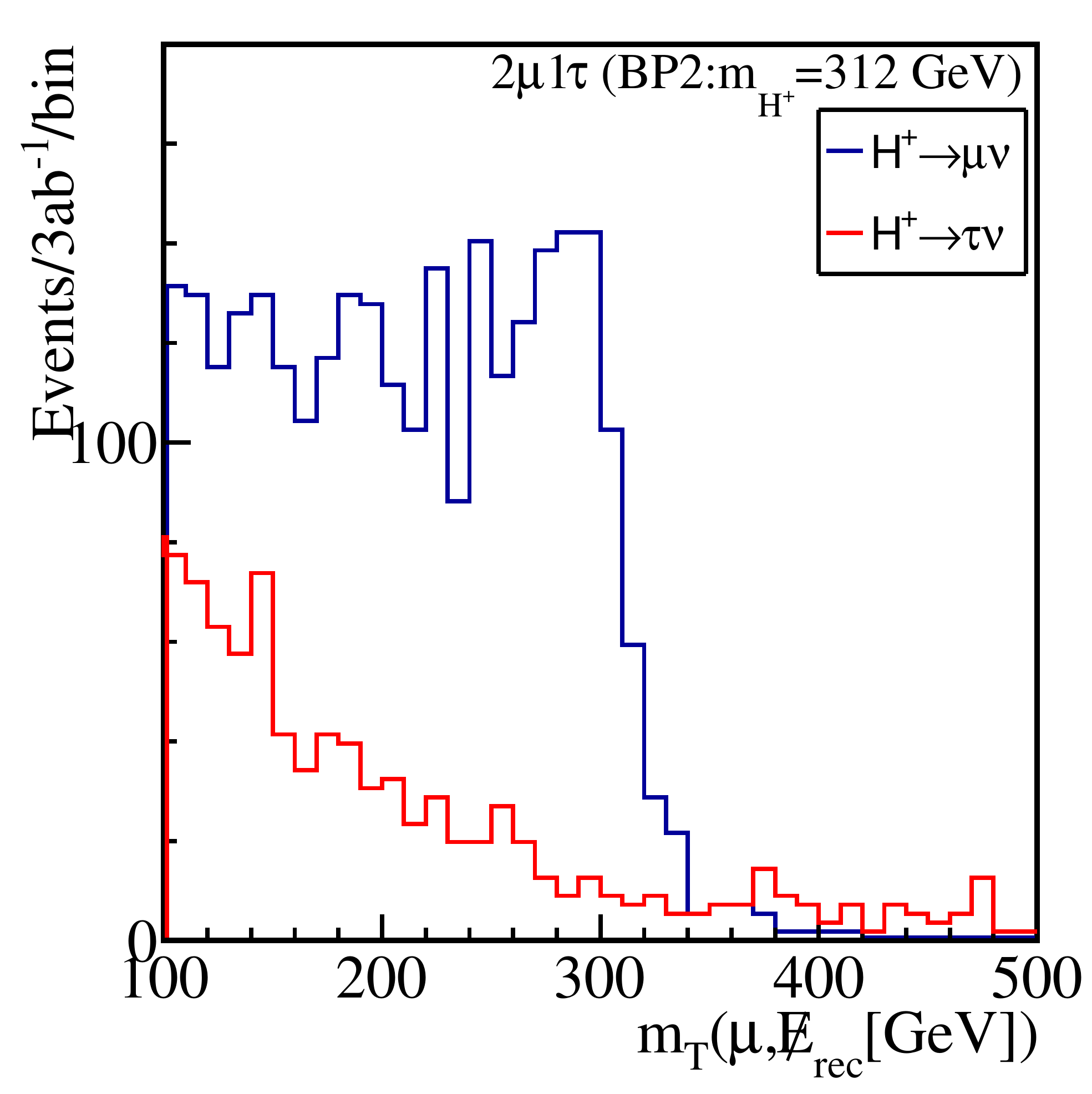}
            \caption{ The $m_{T,\tau}^{\min}$ distribution for $1\mu 2\tau$ mode (left panels) 
            and the $m_{T,\mu}^{\min}$ distribution for $2\mu1\tau$ mode (right panels).
            The upper (lower) panels are for BP1 (BP2).}
    \label{3lep} 
  \end{center}
\end{figure}
%%%%%%%%%
The first difficulty can be partly solved by using the information obtained in the previous 4 lepton mode.
We will take the well-known $m_T$ variable to address the second difficulty, where we also adopt 
the collinear approximation for the $\tau$ momentum reconstruction. It will be a good approximation 
for the taus coming from the heavy resonance. For the preselection, we require one isolated muon 
and two $\tau$-tagged jets ($1\mu2\tau$-mode), 
or two isolated muons and one $\tau$-tagged jet ($2\mu1\tau$-mode), 
with $|\mathbf{p}_{T,\tau}^{\rm vis}|  > 30$~GeV, and $E\!\!\!/_T > 10$~GeV.

For $1\mu2\tau$-mode, relying on the collinear approximation, we define the reconstructed tau 
momenta for $i=1,2$ as 
\begin{eqnarray}
\mathbf{p}_{\tau_i}^{\rm rec} = (1+c_{\tau_i\phi}) \mathbf{p}_{\tau_i}^{\rm vis}, \ \ \ (c_{\tau_i\phi} > 0).
\end{eqnarray}
We first determine the four possible $c_{\tau_i\phi}$'s satisfying the condition 
$m_{\mu \tau_i^{\rm rec}}^2 =(p_{\mu} + p_{\tau_i}^{\rm rec})^2= m_\phi^2$,
corresponding to the four possible hypothesis for the intermediate $\phi=A$ and $H$ and the either $\tau_i$ ($i=1,2$) is from the $\phi$ decay.
For each hypothesis, we define the subtracted missing momentum $\mathbf{p}\!\!\!/_{T, \tau_i\phi}^{\rm sub} = \mathbf{p}\!\!\!/_T - c_{\tau_i\phi}  \mathbf{p}_{T,\tau}^{\rm vis}$, 
and compute the transverse mass $m_{T,\tau_i\phi}=m_{T} ( \mathbf{p}_{\tau_{i^\prime}}^{\rm vis}, \mathbf{p}\!\!\!/_{T, \tau_i\phi}^{\rm sub})$, where $i^\prime = 2, 1$ for $i= 1, 2$, respectively.
Finally we take the minimum of the four $m_{T,\tau_i\phi}$'s as, 
\begin{eqnarray}
m_{T,\tau}^{\rm min} =\min (m_{T,\tau_1A}, m_{T,\tau_1H}, m_{T,\tau_2A}, m_{T,\tau_2H}).
\end{eqnarray}

For $2\mu1\tau$-mode, we similarly define the reconstructed tau momentum 
\begin{eqnarray}
\mathbf{p}_{\tau}^{\rm rec} = (1+c_{\mu_i\phi}) \mathbf{p}_{\tau}^{\rm vis}, \ \ \ (c_{\mu_i\phi} > 0).
\end{eqnarray}
We first determine four possible $c_{\mu_i\phi}$ corresponding to the four possible hypothesis, $m_{\mu_i \tau^{\rm rec}}^2 =(p_{\mu_i} + p_\tau^{\rm rec})^2= m_\phi^2$, where $i=1,2$ and $\phi=A, H$.
For each hypothesis, we compute $m_{T,\mu_i\phi}=m_{T} ( \mathbf{p}_{\mu_{i^\prime}}, \mathbf{p}\!\!\!/_{T, \mu_i\phi}^{\rm sub})$ based on the corresponding subtracted missing momentum $\mathbf{p}\!\!\!/_{T, \mu_i\phi}^{\rm sub} = \mathbf{p}\!\!\!/_T - c_{\mu_i\phi}  \mathbf{p}_{T,\tau}^{\rm vis}$, and $i^\prime = 2, 1$ for $i= 1, 2$, respectively.
Finally we take the minimum of the four $m_{T,\mu_i\phi}$, 
\begin{eqnarray}
m_{T,\mu}^{\rm min} =\min (m_{T,\mu_1A}, m_{T,\mu_1H}, m_{T,\mu_2A}, m_{T,\mu_2H}).
\end{eqnarray}

We show the $m_{T,\tau}^{\rm min}$ and $m_{T,\mu}^{\rm min}$ distributions on the left and right panels
in Fig.~\ref{3lep}, respectively. The upper two panels are for BP1, while the lower two panels are for BP2.
Note that in this procedure, we have used the $m_A$ and $m_H$ values assuming already known from the 4 lepton analysis.

By definition, $m_{T,\mu}^{\rm min}$ ($m_{T,\tau}^{\rm min}$) should be smaller than 
the $m_{T,\mu}$ ($m_{T,\tau}$) with the correct hypothesis, therefore, the endpoint of 
the $m_{T,\mu}^{\rm min}$ distribution should be bounded by the $m_{H^{\pm}}$. We 
see from the plots that the $m_{H^{\pm}}$ information can be extracted from the 
endpoint of these distributions. For all panels, we assume $|\rho_e^{\tau\mu}|=|\rho_e^{\mu\tau}|$, 
therefore, $BR(H^+ \to \tau^+ \nu)=BR(H^+ \to \mu^+ \nu)=50~\%$, and the red lines 
show the contributions from $H^+ \to \tau^+ \nu$ while blue lines show the $H^+ \to \mu^+ \nu$ 
contributions. For the different branching ratio setup, the results would be easily estimated 
by rescaling each contribution. Hence, we can determine the branching ratios from 
the signal ratio of the two modes. Note that there are finite contributions to the $2\mu 1\tau$ 
modes even from the $H^+ \to \tau^+ \nu$ contributions, which are due to the leptonic 
tau decays. For those contributions, $m_{T,\mu}^{\rm min}$ distributions exhibit the same 
endpoint although not steep. On the other hand, there are essentially no $H^+ \to \mu^+ \nu$ 
contributions to $1\mu2\tau$ mode as expected.

%%%%%%%%%%%%%%%%%%%%%%%%%%%%%%%%%%%%
%%%%%%%%%%%%%%%%%%%%%%%%%%%%%%%%%%%%
\subsubsection{2 lepton modes}
\begin{figure}[t]
  \begin{center}
    \includegraphics[width=5cm]{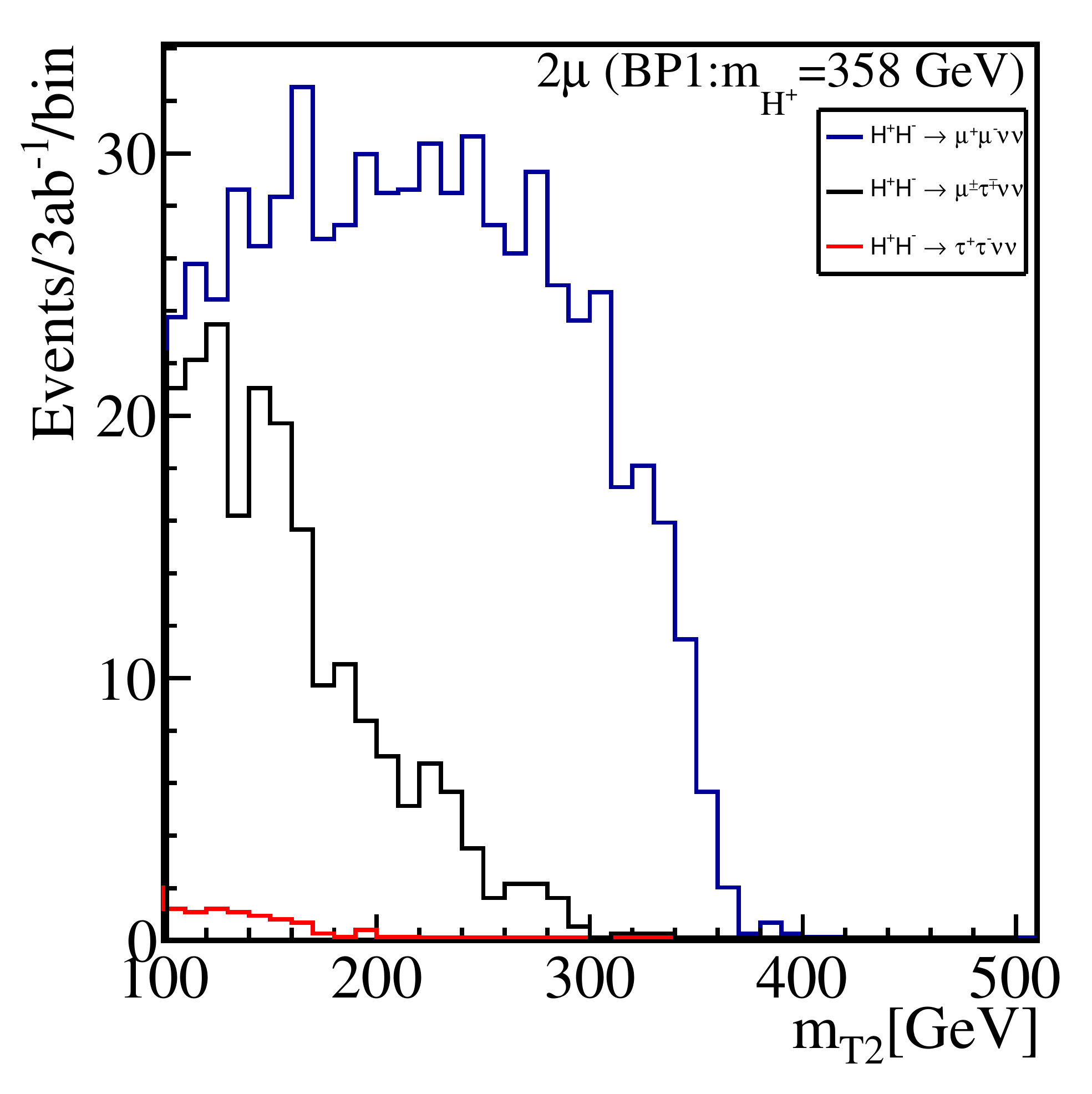}
    \includegraphics[width=5cm]{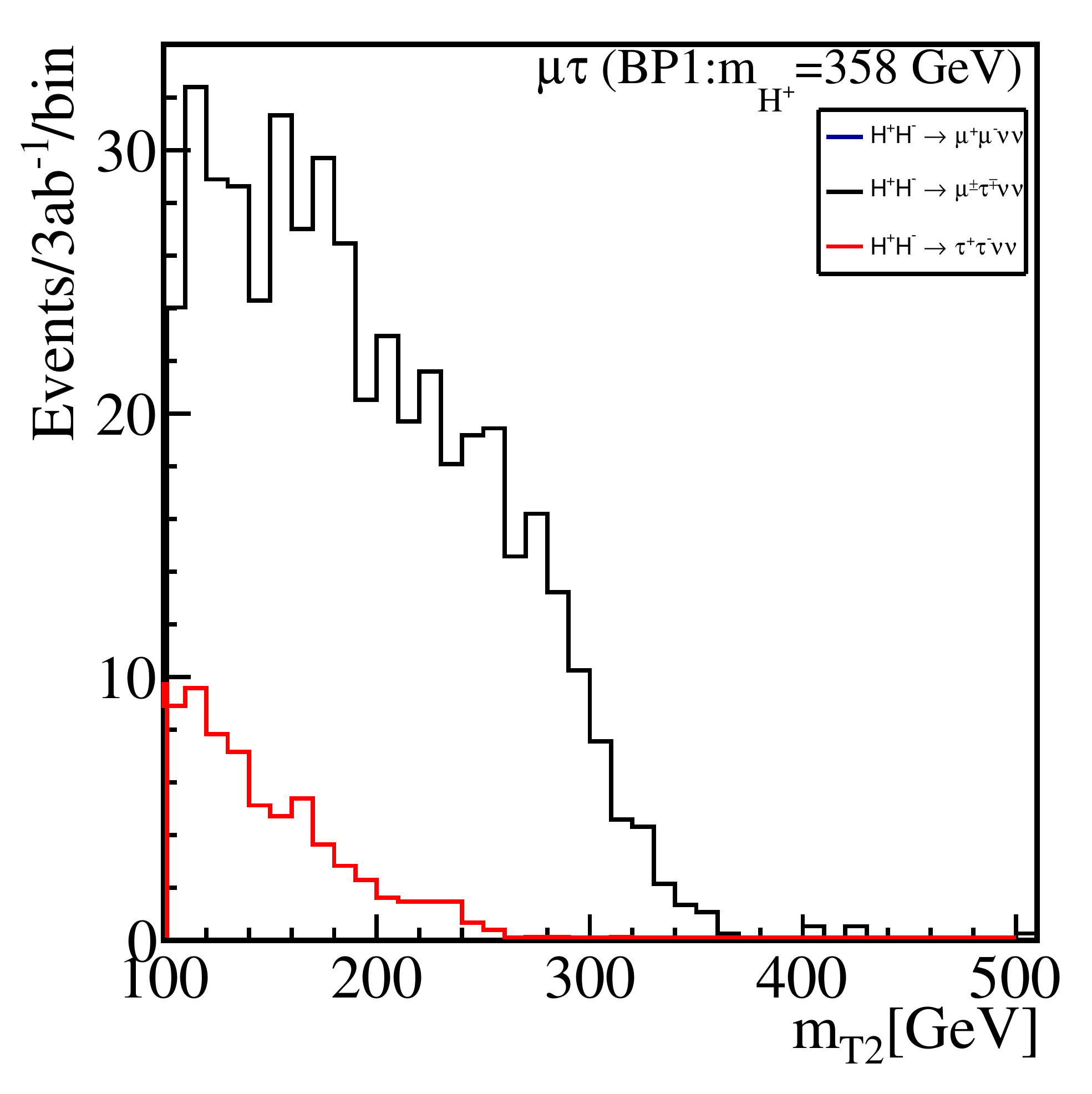}
    \includegraphics[width=5cm]{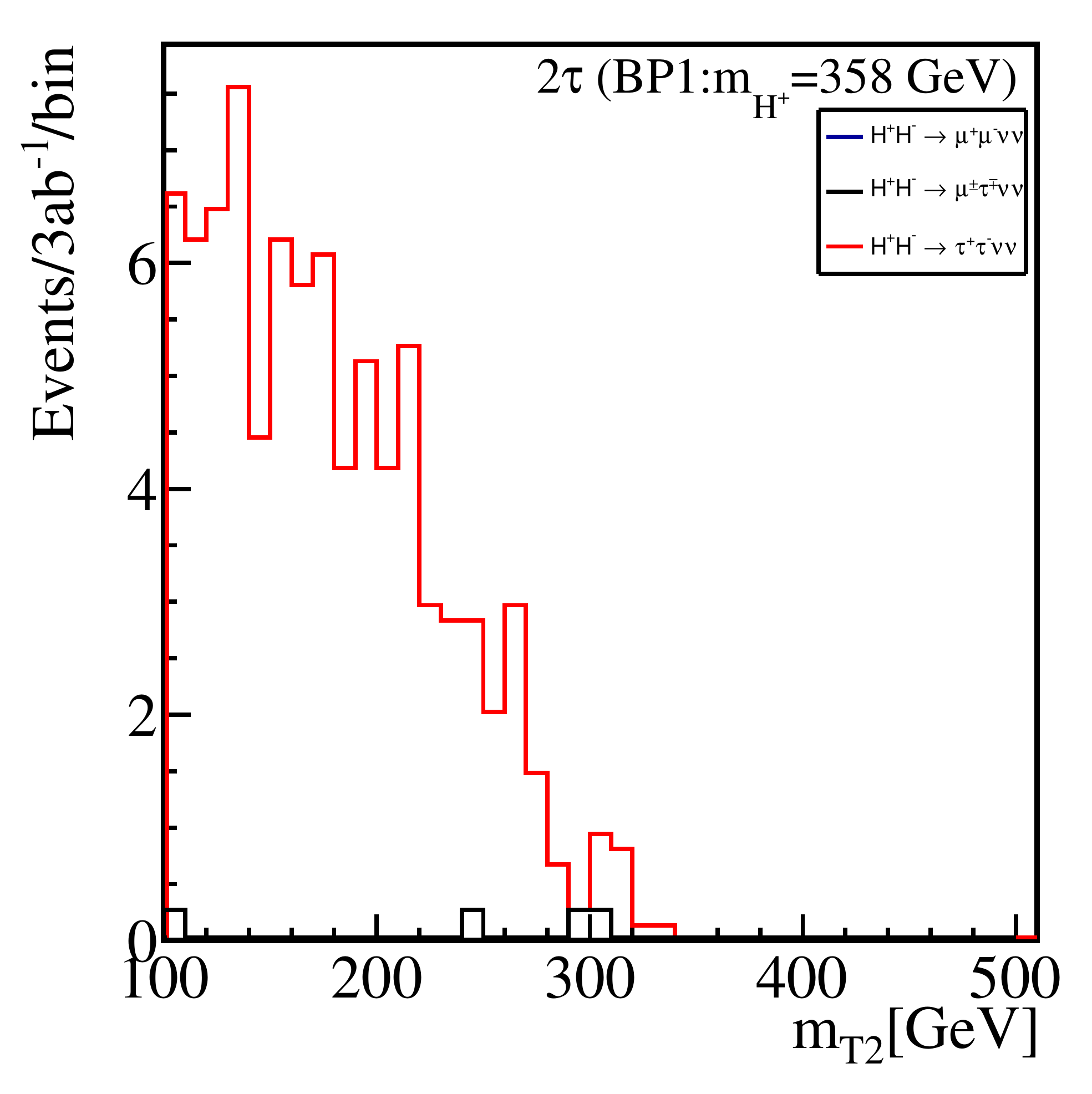}
            \caption{The $m_{T2}$ distribution for BP1 in the $2\mu$ (left), $\mu\tau$ (center), and $2\tau$ (right) modes are shown.}
    \label{2lep}
  \end{center}
\end{figure}

%%%%%%%%%%%%%%%%%%%%%%%%%%%%%%%%%%%%
%%%%%%%%%%%%%%%%%%%%%%%%%%%%%%%%%%%%
Further, we consider the 2-lepton modes from $H^+H^-$ production.
We require the events has $E\!\!\!/_T>100$~GeV for the preselection.
Depending on the branching ratio $r = BR(H^+ \to \tau^+\nu)$, 
$2\mu$, $\mu\tau$, and $2\tau$ modes would be obtained with the fraction of $(1-r)^2:2r(1-r):r^2$, respectively.
Fig.~\ref{2lep} shows the $m_{T2}$ distributions for 
the each category of the events for BP1, where the $m_{T2}$ is defined as follows, 
\begin{eqnarray}
m_{T2}(\mathbf{p}_{\ell_1}, \mathbf{p}_{\ell_2},\mathbf{p}\!\!\!/_T) 
= \min_{\mathbf{p}\!\!/_T=\mathbf{p}\!\!/_{T,1}+\mathbf{p}\!\!/_{T,2}}
\{\max [m_{T}(\mathbf{p}_{\ell_1},\mathbf{p}\!\!\!/_{T,1}),
m_{T}(\mathbf{p}_{\ell_2},\mathbf{p}\!\!\!/_{T,2}) ]\},
\end{eqnarray}
and each $\ell_i = \mu$ or $\tau^{\rm vis}$.
On all panels, $r=0.5$ is assumed, and the blue, black and red lines show the contributions from both $H^\pm$ decay into $\mu$'s ($\mu^+\mu^-\nu\nu$, 25~\%), 
each into $\mu$ and $\tau$ respectively ($\mu^\pm\tau^\mp\nu\nu$, 50~\%), and both into $\tau$'s ($\tau^+\tau^-\nu\nu$, 25~\%), respectively.

%%%%%%%%%
%%%%%%%%%
For $2\mu$-mode, $m_{T2}$ distribution from $\mu^+\mu^-\nu\nu$ mode has a clear endpoint at $m_{H^{\pm}}$, therefore, we can determine the 
$H^\pm$ mass, as long as $BR(H^\pm \to \mu^\pm \nu)$ is sizable. 
For $\mu\tau$-mode, the $m_{T2}$ endpoint is rather smeared due to the escaping missing momentum by the extra neutrinos.
The main contributions to this mode is from the events where each $H^\pm$ decay into $\mu$ and $\tau$ respectively. 
For $2\tau$-mode, the endpoint is further smeared and locates at the lower value.
For $2\mu$-mode, $\mu\tau$-mode, and $2\tau$-mode, roughly 50~\%, 18~\%, and 7~\% of $\mu^+\mu^-\nu\nu$,
$\tau^\pm\mu^\mp\nu\nu$, and $\tau^+\tau^-\nu\nu$ events remain after requiring $m_{T2}>100$~GeV, respectively.
These numbers are understood due to the hadronic $\tau$ branching ratio, the $\tau$-tagging efficiency, and further cancelation of the missing momentum by the extra neutrinos in the $\tau$ decays. Using the relative ratio among the numbers observed in these three modes, we can in principle access the branching ratio information as in the 3 lepton mode shown in the previous section.

%%%%%%%%%%%%%%%%%%%%%%%%%%%%%%%%%%%%
%%%%%%%%%%%%%%%%%%%%%%%%%%%%%%%%%%%%
%%%%%%%%%%%%%%%%%%%%%%%%%%%%%%%%%%%%
\section{Discussion}
\label{DCS}
%%%%%%%%%%%%%%%%%%%%%%%%%%%%%%%%%%%%
%%%%%%%%%%%%%%%%%%%%%%%%%%%%%%%%%%%%
%%%%%%%%%%%%%%%%%%%%%%%%%%%%%%%%%%%%
In this section, we explore the possible parameter space for the other Yukawa elements when  the product $\rho_{e}^{\mu\tau}\rho_{e}^{\tau\mu}$ is sizable to explain $\delta a_\mu$ 
and evaluate the effect to the LHC signatures. In general, if the other elements are sizable, BR($\phi\to\mu\tau$) 
will be diluted, and the multi-lepton signatures considered in the previous section 
would be reduced. We evaluate how large the dilution effects could be by using the 
parameters consistent with the experimental constraints. We estimate it by adding 
each element to the BP1 as a reference. First of all, since Yukawa elements for the 
1st and 2nd generation quarks are stringently constrained by the flavor and 
collider experiments, their allowed value is extremely small and would not practically 
reduce the signals. Hence, it leaves our focus on those for the third generations: 
$\rho_e^{\tau\tau},~\rho_u^{tt},~\rho_u^{tc}$,~$\rho_u^{ct}$ and $\rho_d^{bb}$.

First, $\tau\to\mu\gamma$ would be induced through a 1-loop diagram proportional 
to the $\rho_e^{\tau\tau}$ and through the 2-loop Barr-Zee diagram proportional to the 
$\rho_u^{tt}$~\cite{Omura:2015nja}.  The observed $BR(\tau\to\mu\gamma)$ 
sets the stringent upper limits as $|\rho_u^{tt}|<0.05$ and $|\rho_e^{\tau\tau}| < 0.06$.
Let us give a comment on $\rho_e^{\mu\mu}$. The products of $\rho_e^{\tau\mu}\rho_e^{\mu\mu}$ and $\rho_e^{\mu\tau}\rho_e^{\mu\mu}$ induce a dangerous contribution to $\tau\to3\mu $ at tree level mediated by $H$ and $A$.
The observed upper limit on $BR(\tau\to3\mu)$ constrains as $|\rho_e^{\mu\mu}| < O(10^{-3})$\cite{Omura:2015xcg}.
Next, $|\rho_u^{tc}| < 0.11$ is obtained by the light lepton universality in the 
semi-leptonic decays of the $B$ meson, $B\to Dl\nu$, where $l=\mu,e$~\cite{Iguro:2017ysu}. 
The $\epsilon_K$ measurements provide a severe constraint as $|\rho_u^{ct}|<0.04$~\cite{IO2}.
For $|\rho_u^{tc}|$, the current LHC data have the potential to set the most stringent constraint
through the $\tau \nu$ and the $\mu\nu$ resonance searches~\cite{Iguro:2018fni,Khachatryan:2016jww}.
%(when the charged Higgs mass is heavier than $400$~GeV ) 
There is, however, no dedicated study available to target it. 
The only available LHC search to constrain the parameter  $\rho_u^{tc}$ is the one for the same-sign di-tops, and it sets the significantly weaker upper limit: $|\rho_u^{tc}| < 0.7$~\cite{Hou:2018zmg}.
Finally, $|\rho_d^{bb}| < 0.22$ is obtained by the flavor observables including $BR(B\to\mu\nu)$ 
and $BR(B\to\tau\nu)$~\cite{Prim}.  For $\rho_d^{bb}$, the constraints from the collider 
experiments are discussed in Type~II 2HDM, and those are applicable to our setup 
although the constraints are weaker than the ones from the flavor experiments.
In summary, $\rho_d^{bb}$ is the element allowed to take the largest value among the five elements 
listed above. When $|\rho^{\tau\mu}_e|=|\rho^{\mu\tau}_e|=0.3$, $|\rho_d^{bb}| < 0.22$ implies 
the $BR(\phi\to bb) = 3|\rho_d^{bb}|^2/(|\rho^{\tau\mu}_e|^2 +|\rho^{\mu\tau}_e|^2+3|\rho_d^{bb}|^2) < 0.3$, 
and therefore, phenomenologically our multi-lepton signals can decrease by a third at most.

For a certain fixed value of $m_A$, the larger $\Delta_{H-A}$ is assigned, the smaller product of $|\rho_e^{\mu\tau} \rho_e^{\tau\mu}| \propto \Delta_{H-A}^{-1}$ is required to obtain the same $\delta a_\mu$.
The larger $|\rho_e^{\mu\tau} \rho_e^{\tau\mu}|$ faces
the more stringent constraints on the other Yukawa couplings; for example, 
$BR(B\to\mu\nu)$ constrains the product  $\rho_e^{\tau\mu} \rho_d^{bb}$, therefore,
the upper bound on $|\rho_d^{bb}|$ scales $\propto 1/\rho_e^{\tau\mu}$. 
As a result, a scaling $BR(\phi\to bb) \propto 1/|\rho_e^{\tau\mu}|^4$ is obtained, where we assumed 
$|\rho^{\tau\mu}_e|=|\rho^{\mu\tau}_e|$.
Therefore, as the $\Delta_{H-A}$ decreases, that corresponds to increasing $|\rho_e^{\mu\tau} \rho_e^{\tau\mu}|$, the dilution effect quickly vanishes.

When $\Delta_{H-A}$ becomes larger than $W$ and $Z$ boson masses, the decays of
$H \to W^\pm H^\mp$ and $H \to A Z$ are kinematically allowed, 
and as a result $BR(H\to\mu\tau)$ decreases significantly.
For the former mode, the leptonic branching ratio would be reduced, 
while for the latter mode the subsequent decay of $A$ like $H\to A Z\to \tau\mu Z$ 
would again contributes to the multi-lepton signatures. Additional $Z$ can even 
provide extra leptons and it would result in a more characteristic
signature.
\medskip

As we have demonstrated in the previous section, $m_H$, $m_A$, $m_{H^\pm}$, and 
the ratio of $|\rho_e^{\mu\tau}/\rho_e^{\tau\mu}|$ can be reconstructed at the LHC 
among the minimal set of the five parameters. Although we have not shown explicitly, 
the ratio is also accessible by measuring the chirality of the leptons from the $\phi$ decays.
On the other hand, the absolute size of the product $\rho_e^{\mu\tau} \rho_e^{\tau\mu}$ 
would be insensitive to the LHC signatures and difficult to determine. For this purpose, 
the existence of the other Yukawa elements would be helpful.  For example, 
a finite $\rho_d^{bb}$ opens another production mode $b\bar{b}\phi$, 
which would contribute to another source of the multi-lepton events. 
If we can identify the $b\bar{b}\phi$ production events, we can independently 
access the information on $\rho_d^{bb}$ and the ratio $|\rho_d^{bb}/\rho_e^{\tau\mu}|$,
which means the absolute value of the $|\rho_e^{\tau\mu}\rho_e^{\mu\tau}|$ is measurable.
Similarly, when $H \to W^\pm H^\mp$ and $H \to AZ$ open we can access it via the relative size of 
those modes against the $H \to \mu^\pm\tau^\mp$ mode since partial widths of those modes are controlled 
by the weak gauge coupling.

%%%%%%%%%%%%%%%%%%%%%%%%%%%%%%%%%%%%
%%%%%%%%%%%%%%%%%%%%%%%%%%%%%%%%%%%%
\section{Summary}
Motivated by the discrepancy between the experimentally measured value and the SM prediction 
of the muon anomalous magnetic moment, we consider the 2HDM with the lepton flavor violating 
Yukawa couplings $\rho_e^{\mu\tau}$ and $\rho_e^{\tau\mu}$.  We show the preferred heavy 
Higgs masses are of ${\cal O}(100)$~GeV and limited below $\sim 700$~GeV requiring the 
perturbativity of the couplings. 

We have pointed out that this scenario predicts the very characteristic multi-lepton signatures from the pair production of the heavy resonances $HA$, $H^\pm\phi$, and $H^+H^-$ via the electroweak production.
Among them, the 4 lepton signatures $\mu^+\mu^-\tau^+\tau^-$, and especially $\mu^\pm\mu^\pm\tau^\mp\tau^\mp$ would be very distinctive. We estimate that the current data accumulated at the LHC are enough 
sensitive to a part of the parameter region in this scenario, therefore, the experimental searches 
targeting those signatures are strongly desired.

As demonstrated in Sec.~\ref{sec:lhc}, once enough data are accumulated, 
reconstructing their mass spectrum would be possible using the reconstructed 
invariant masses, $m_T$, and $m_{T2}$ distributions.  For the momentum reconstruction 
of taus, the collinear approximation plays an important role, which would be a good approximation 
for a boosted taus from the decay of such heavy particles. We estimate the resolution 
of the reconstructed mass difference between $A$ and $H$, $\Delta_{H-A}$, and show 
that resolving $\Delta_{H-A} \sim {\cal O}(20)$~GeV would be achievable.
Note that it is easier to accommodate a sizable $\delta a_\mu$ contribution with the larger $\Delta_{H-A}$, and our study shows it promising to identify  the existence of two resonances in most 
of the relevant parameter region.

Furthermore, we can measure the ratio of the couplings $|\rho_e^{\mu\tau}|$ and $|\rho_e^{\tau\mu}|$ 
via the ratio between $BR(H^\pm\to\mu^\pm\nu)$ and $BR(H^\pm\to\tau^\pm\nu)$, which would be extracted by the ratio among the 3 lepton, and 2 lepton modes. The ratio is also accessible from the chirality of the leptons from heavy extra Higgs decays. More complicated setups including other Yukawa elements and other decay modes would help to determine the absolute size of the couplings $|\rho_e^{\mu\tau}\rho_e^{\tau\mu}|$.
\medskip

Since our signatures rely on the weak interaction, the same analysis at the lepton colliders 
such as the ILC would be performed as long as $\sqrt s$ is large enough, 
where we possibly determine the model parameters more precisely 
in the cleaner environments and using the energy conservation.

%%%%%%%%%%%%%%%%%%%%%%%%%%%%%%%%%%%%
%%%%%%%%%%%%%%%%%%%%%%%%%%%%%%%%%%%%
\section*{Acknowledegments}
The authors also thank Junji Hisano, Kazuhiro Tobe, Tomomi Kawaguchi, Shigeki Hirose, and Makoto Tomoto for valuable discussions. 
%%%%%%%%%%%%%%%%%%%%%%%%%%%%%%%%%%%%
%---------------------------------------------------------------------------
The work of S. I. is supported by Kobayashi-Maskawa Institute for the Origin of Particles and the Universe, Toyoaki scholarship foundation and the Japan Society for the Promotion of Science (JSPS) Research Fellowships for Young Scientists, No. 19J10980.
%---------------------------------------------------------------------------
The work of Y. O. is supported by Grant-in-Aid for Scientific research from the Ministry of Education, Science, Sports, and Culture (MEXT), Japan, No. 19H04614, No. 19H05101 and No. 19K03867.
%---------------------------------------------------------------------------
M. T. is supported in part by the JSPS Grant-in-Aid for Scientific Research
Numbers~16H03991, 16H02176, 18K03611, and 19H04613.
%---------------------------------------------------------------------------

%%%%%%%%%%%%%%%%%%%%%%%%%%%%%%%%%%%%
%%%%%%%%%%%%%%%%%%%%%%%%%%%%%%%%%%%%

\end{document}